\documentclass[%
 reprint,
 amsmath,amssymb,
 aps,
]{revtex4-2}

\usepackage{graphicx}
\usepackage{dcolumn}

\usepackage{bm}
\usepackage{bbm}

\usepackage{comment}  

\usepackage{xcolor} 
\usepackage{multirow}
\usepackage{xfrac}

\usepackage{ulem}

\usepackage{tikz, tkz-euclide, ifthen}
\usetikzlibrary{quotes, angles}
\usepackage{listofitems} 

\definecolor{gold}{rgb}{0.95, 0.69, 0.24}
\definecolor{grey}{rgb}{0.57, 0.57, 0.57}

\pgfset{
	foreach/parallel foreach/.style args={#1in#2via#3}{evaluate=#3 as #1 using {{#2}[#3-1]}},
}

\definecolor{amethyst}{rgb}{0.6, 0.4, 0.8}

\begin{document}
\preprint{APS/123-QED}



\title{Tuning Plasma Frequency of Nested Wire Media}



\author{Denis Sakhno$^{1}$}
\email{denis.sakhno@metalab.ifmo.ru}

\author{Pavel A. Belov$^{1,2}$}%

\affiliation{$^1$School of Physics and Engineering, ITMO University, Kronverksky Pr. 49, 197101, St. Petersburg, Russia}

\affiliation{$^2$School of Engineering, New Uzbekistan University, Movarounnahr str. 1, 100000, Tashkent, Uzbekistan}

\date{\today}

\begin{abstract}

We study nested wire media possessing $C_6$ and $C_4$ rotational symmetries whose plasma frequenc{ies} 
can be controlled through breathing deformation. Numerical simulations reveal tunability exceeding $80\%$ and $60\%$ for the considered $C_6$ and $C_4$ breathing geometries{,} respectively. To describe these structures, we develop an analytical model based on the local field approach within the thin-wire approximation, which confirm{s} 
the high tunability 
{of} the studied 
wire structures. {We also propose an 
approximation for dense wire media.} 
The developed framework is applicable to the general case of nested wire structures -- wire media with an arbitrary parallelogram lattice and multiple wires per unit cell.

\end{abstract}

\maketitle


\section{Introduction}


\textit{Wire media} {are} 
a distinct class of electromagnetic artificial materials composed of metallic rods with different periodicities, electrical connections, and topologies \cite{simovski2012wire, sakhno2026microwave}. These structures {were among the first metamaterials} 
\cite{engheta2006book, capolino2009book01, capolino2009book02, shalaev2010book}{, with first descriptions of wire media in} 
\cite{sievenpiper19963d, pendry1996extremely}. Metamaterials composed of wires possess strong spatial dispersion \cite{landau1984electrodynamics} (strong non-locality) because electric charges can freely propagate throughout the material volume. This non-locality enables several practical applications of wire media, such as 
artificial plasma \cite{silveirinha2009artificial}, 
subwavelength imaging devices \cite{belov2006subwavelength, silveirinha2007imaging}, and 
improving the performance of magnetic resonance imaging devices \cite{guo2024metamaterial}.

A \textit{simple wire medium} is an array of parallel wires oriented along a given direction. This structure is one of the earliest studied wire media and was investigated as a {``}rodded-type artificial dielectric{"} \cite{brown1950design,brown1953artificial,carne1959theory,brown1960artificial, rotman1962plasma} even before the term 
{``}metamaterials" appeared in the scientific literature. This metamaterial suppresses the propagation of TM-polarized electromagnetic waves (with the electric field parallel to the wires) below \textit{the plasma frequency} ($\omega_p$). For TM-waves at frequencies slightly above the plasma frequency, a simple wire medium becomes an epsilon-near-zero (ENZ) medium \cite{burghignoli2008directive, li2013radiation}.

Rectangular lattice{s} of infinite wires {are} 
the most extensively studied wire medi{a} 
\cite{pendry1998low, belov2002dispersion, maslovski2009nonlocal, tyukhtin2011effective}. 
Other lattice types \cite{nicorovici1995photonic, smirnova2002simulation, silveirinha2003efficient, park2009efficient} have received less attention because no significant differences in observable physical effects were expected compared to rectangular lattices.

\textit{Nested wire media} \cite{fernandes2013fano} constitute a more complex class of wire media -- 
periodic structures containing several ($N>1$) wires, possibly of different radii $r_n$, within a single unit cell. Such structures can be viewed as a superposition of $N$ interpenetrating wire sub-lattices, each formed by wires of radius $r_n$. Nested wire structures allow currents within the same unit cell to flow in different directions due to electromagnetic coupling between the sub-lattices. The case of two nested ($N=2$) rectangular lattices of parallel wires was investigated in \cite{fernandes2013fano}, where 
sharp Fano resonances 
{emerged during} 
the scattering of electromagnetic waves by a finite-thickness slab.

In recent years, interest in wire media has 
{rekindled} due to the{ir novel application} 
in cosmology-related experiments for axion searches \cite{dine1983not, wilczek1987two}.
{For this purpose,} the plasma haloscope concept \cite{balafendiev2022resonator, millar2023alpha} was proposed: 
wire-medium-filled resonators provide the required tunability in the microwave range via mechanical tuning {of the plasma frequency} (by changing the periodicity of the medium), 
while keeping the resonator volume unchanged and suitable for operation at cryogenic temperatures.

The need for mechanical tunability of wire media has led to the development of different tuning approaches. These range from nested geometries \cite{kowitt2023tunable, sakhno2025honeycomb} 
to designs based on rotating thick wires with different cross-sections \cite{balafendiev2025tunable} and 
auxetic-inspired deformable designs \cite{bae2023tunable}.
Despite the extensive study of tunable geometries based on rectangular lattices, 
triangular lattice{s} \cite{sakhno2025honeycomb, balafendiev2025tunable} offer 
significant tunability potential, which motivates further investigations of other lattice types{,} at least to {enhance} 
flexibility in the plasma frequency tuning.

\begin{figure*}[t]
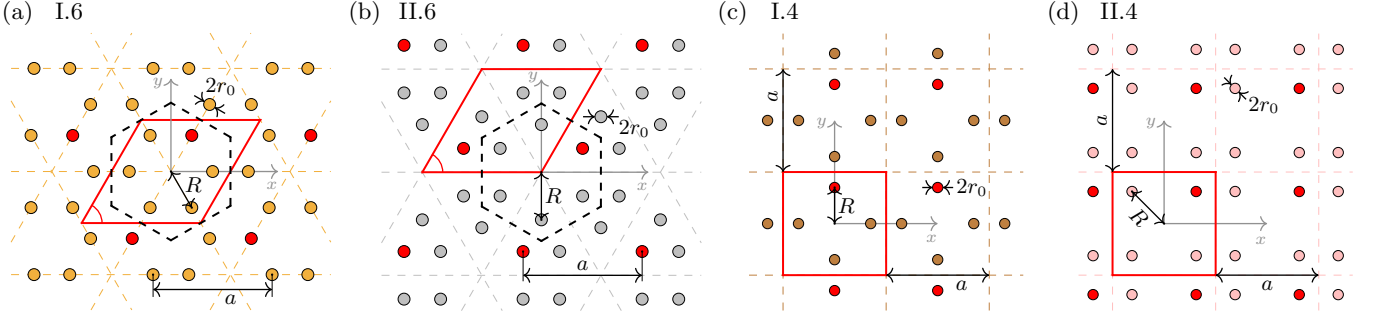

    \begin{minipage}{0.26\linewidth}
		\center{ 
			\input{fig/hexagon/03_geom_hex_2ribs.tikz}
		}
	\end{minipage}
    \put(-133, 60){(a)}
    \put(-113, 60){I.6}
    \hfill
	\begin{minipage}{0.26\linewidth}
		\center{
            \input{fig/hexagon/04_geom_hex_2verts.tikz}
		}
	\end{minipage}
    \put(-142, 60){(b)}
    \put(-123, 60){II.6}
    \hfill
	\begin{minipage}{0.21\linewidth}
		\center{
            \definecolor{gold}{rgb}{0.95, 0.69, 0.24}
\definecolor{grey}{rgb}{0.57, 0.57, 0.57}

\begin{tikzpicture}[scale=0.90, transform shape]
	\def \c{1.75 * sin(60)}  
	\def \a{1.75 * sin(60)}  
	\def \ang{90}  
    \def \rot {0}
	\def \RtoA {0.35}

    \readlist*\wCols{brown, red, brown, brown}
    
	\def \shiftx {0}
	\def \shifty {0}
	
	\coordinate(A1) at ({0 + \shiftx}, {0 + \shifty}); 
	\coordinate(A2) at ({\a + \shiftx}, {0 + \shifty}); 
	\coordinate(A3) at ({\a + \c * cos(\ang) + \shiftx}, {\c * sin(\ang) + \shifty}); 
	\coordinate(A4) at ({\c * cos(\ang) + \shiftx}, {\c * sin(\ang) + \shifty}); 
	\def \verts {A1, A2, A3, A4}
	
    
	\def \rw{0.05 * \a}  
	\def \wirescol{brown}  
    
	
	\def \numa {2}  
	\def \numc {2}  
	\def \delta {0.5}  
    \def \deltaY {0.6 * sin(60)}
	\def \linesWidth {0.25pt}
    
	\foreach [evaluate={
		\linex = \c * cos(\ang) * \numc ;
		\liney = \c * sin(\ang) * \numc 
	}] \ia in {0, ..., \numa} {
	
		\coordinate(diag1) at ({ \ia * \a - \deltaY * cos(\ang) }, {- \deltaY * sin(\ang) });
		\coordinate(diag2) at ({ \ia * \a + \linex + \deltaY * cos(\ang) }, { \liney + \deltaY * sin(\ang) });
		
		\draw[dashed, line width=\linesWidth, \wirescol](diag1) -- (diag2);
	}
    
	\foreach [evaluate={
		\rowy = \c * sin(\ang) * \ic
	}] \ic in {0, ..., \numc} {
		
		\coordinate(left) at ({ -\delta + \c * cos(\ang) }, { \rowy });
		\coordinate(right) at ({ 2 * \a + \c * cos(\ang) + \delta }, { \rowy });
		
		\draw[dashed, line width=\linesWidth, \wirescol](left) -- (right);
	}

    \coordinate(left) at ({ -\delta + \c * cos(\ang) }, { \c * sin(\ang) * 2.61 });
    \coordinate(right) at ({ 2 * \a + \c * cos(\ang) + \delta }, { \c * sin(\ang) * 2.61 });
    
    \draw[dashed, line width=\linesWidth, white](left) -- (right);

    
	\def \refX {0.5 * \a + 0.5 * \c * cos(\ang)}
	\def \refY {0.5 * \c * sin(\ang)}
	
	\draw[->, line width=0.5pt, grey] ({\refX}, {\refY}) -- ({\refX + \a * 1}, {\refY}) node[left, xshift=3pt, yshift=-6pt] {\scriptsize $x$};
	\draw[->, line width=0.5pt, grey] ({\refX}, {\refY}) -- ({\refX}, {\refY + \c * sin(\ang) * 1}) node[left, , xshift=0pt, yshift=-1pt] {\scriptsize $y$};
    
	\draw[solid, line width=0.75pt, red](A1) -- (A2) -- (A3) -- (A4)--(A1);
	\foreach \vert in \verts {
		\filldraw [red] (\vert) circle (0.25pt); 
	}

    


    \foreach \ia in {0, 1} {
        \foreach \ic in {0, 1} {
            \def \centerX {\a / 2 + \a * \ia};
            \def \centerY {\c / 2 + \c * \ic};
            \foreach \i in {0, ..., 3} {
                \coordinate(this) at (
                    { \centerX + \RtoA * \a * cos(\ang * \i + \rot)}, 
                    { \centerY + \RtoA * \a * sin(\ang * \i + \rot)}
                );
                \filldraw [black, fill={\wCols[\i + 1]}, line width=0.25pt] (this) circle (\rw);
            }
        }
    }

    \foreach \ia in {0, 1} {
        \def \ic {-1};
        \def \centerX {\a / 2 + \a * \ia};
        \def \centerY {\c / 2 + \c * \ic};
        \foreach \i in {1} {
            \coordinate(this) at (
                { \centerX + \RtoA * \a * cos(\ang * \i + \rot)}, 
                { \centerY + \RtoA * \a * sin(\ang * \i + \rot)}
            );
            \filldraw [black, fill={\wCols[\i + 1]}, line width=0.25pt] (this) circle (\rw);
        }
    }

    \foreach \ia in {0, 1} {
        \def \ic {2};
        \def \centerX {\a / 2 + \a * \ia};
        \def \centerY {\c / 2 + \c * \ic};
        \foreach \i in {3} {
            \coordinate(this) at (
                { \centerX + \RtoA * \a * cos(\ang * \i + \rot)}, 
                { \centerY + \RtoA * \a * sin(\ang * \i + \rot)}
            );
            \filldraw [black, fill={\wCols[\i + 1]}, line width=0.25pt] (this) circle (\rw);
        }
    }

    \foreach \ic in {0, 1} {
        \def \ia {-1};
        \def \centerX {\a / 2 + \a * \ia};
        \def \centerY {\c / 2 + \c * \ic};
        \foreach \i in {0} {
            \coordinate(this) at (
                { \centerX + \RtoA * \a * cos(\ang * \i + \rot)}, 
                { \centerY + \RtoA * \a * sin(\ang * \i + \rot)}
            );
            \filldraw [black, fill={\wCols[\i + 1]}, line width=0.25pt] (this) circle (\rw);
        }
    }

    \foreach \ic in {0, 1} {
        \def \ia {2};
        \def \centerX {\a / 2 + \a * \ia};
        \def \centerY {\c / 2 + \c * \ic};
        \foreach \i in {2} {
            \coordinate(this) at (
                { \centerX + \RtoA * \a * cos(\ang * \i + \rot)}, 
                { \centerY + \RtoA * \a * sin(\ang * \i + \rot)}
            );
            \filldraw [black, fill={\wCols[\i + 1]}, line width=0.25pt] (this) circle (\rw);
        }
    }
    
	
	
	\def \circX {\a / 2 + 1 * \a + \RtoA * \a * cos(1 * \ang + \rot) + 0 * \c * cos(\ang)}
	\def \circY {\c / 2 + \RtoA * \a * sin(1 * \ang + \rot) + 0 * \c * sin(\ang)}
	\def \angArr {\ang + 90}

	\draw[->, black, line width=0.5pt]({\circX-3*\rw *cos(\angArr)}, {\circY-3*\rw * sin(\angArr)}) -- ({\circX-\rw*cos(\angArr)}, {\circY-\rw*sin(\angArr)}) node[midway,xshift=10pt, yshift=0pt] {$2r_0$};
	\draw[->, black, line width=0.5pt]({\circX+3*\rw *cos(\angArr)}, {\circY+3*\rw * sin(\angArr)}) -- ({\circX+\rw*cos(\angArr)}, {\circY+\rw*sin(\angArr)});
	
    \def \deltaY {2.5 * \rw}
    
	\coordinate (aLeft) at (
        {\a}, {0}
    );
	\coordinate (aRight) at (
        {2 * \a}, {0}
    );
	\draw[<->, line width=0.5pt, black] (aLeft) -- (aRight) node[midway, below, sloped, xshift=10pt, yshift=1pt] {$a$};

    \coordinate (aLow) at (
        {0}, {\c}
    );
	\coordinate (aUp) at (
        {0}, {2 * \c}
    );
	\draw[<->, line width=0.5pt, black] (aLow) -- (aUp) node[midway, above, sloped, xshift=10pt, yshift=-1pt] {$a$};

    \def \deltaY {2.5 * \rw}
    
	\coordinate (aLeft) at (
        {\a / 2}, 
        {\c / 2}
    );
	\coordinate (aRight) at (
        {\a / 2 }, 
        {\c / 2 + \RtoA * \a}
    );
	\draw[<->, line width=0.5pt, black] (aLeft) -- (aRight) node[right, xshift=-2pt, yshift=-7.5pt] {$R$};

    \filldraw [black, fill=black, line width=0.25pt] (aLeft) circle (0.25pt);
    \filldraw [black, fill=black, line width=0.25pt] (aRight) circle (0.25pt);
	
\end{tikzpicture}
		}
	\end{minipage}
    \put(-117, 60){(c)}
    \put(-97, 60){I.4}
    \hfill
    \hspace{7pt}
	\begin{minipage}{0.21\linewidth}
		\center{
            \definecolor{gold}{rgb}{0.95, 0.69, 0.24}
\definecolor{grey}{rgb}{0.57, 0.57, 0.57}

\begin{tikzpicture}[scale=0.90, transform shape]
	\def \c{1.75 * sin(60)}  
	\def \a{1.75 * sin(60)}  
	\def \ang{90}  
    \def \rot {45}
	\def \RtoA {0.44}

    \readlist*\wCols{red, pink, pink, pink}
    
	\def \shiftx {0}
	\def \shifty {0}
	
	\coordinate(A1) at ({0 + \shiftx}, {0 + \shifty}); 
	\coordinate(A2) at ({\a + \shiftx}, {0 + \shifty}); 
	\coordinate(A3) at ({\a + \c * cos(\ang) + \shiftx}, {\c * sin(\ang) + \shifty}); 
	\coordinate(A4) at ({\c * cos(\ang) + \shiftx}, {\c * sin(\ang) + \shifty}); 
	\def \verts {A1, A2, A3, A4}
	
    
	\def \rw{0.05 * \a}  
	\def \wirescol{pink}  
    
	
	\def \numa {2}  
	\def \numc {2}  
	\def \delta {0.5}  
    \def \deltaY {0.6 * sin(60)}
	\def \linesWidth {0.25pt}
    
	\foreach [evaluate={
		\linex = \c * cos(\ang) * \numc ;
		\liney = \c * sin(\ang) * \numc 
	}] \ia in {0, ..., \numa} {
	
		\coordinate(diag1) at ({ \ia * \a - \deltaY * cos(\ang) }, {- \deltaY * sin(\ang) });
		\coordinate(diag2) at ({ \ia * \a + \linex + \deltaY * cos(\ang) }, { \liney + \deltaY * sin(\ang) });
		
		\draw[dashed, line width=\linesWidth, \wirescol](diag1) -- (diag2);
	}
    
	\foreach [evaluate={
		\rowy = \c * sin(\ang) * \ic
	}] \ic in {0, ..., \numc} {
		
		\coordinate(left) at ({ -\delta + \c * cos(\ang) }, { \rowy });
		\coordinate(right) at ({ 2 * \a + \c * cos(\ang) + \delta }, { \rowy });
		
		\draw[dashed, line width=\linesWidth, \wirescol](left) -- (right);
	}

    \coordinate(left) at ({ -\delta + \c * cos(\ang) }, { \c * sin(\ang) * 2.61 });
    \coordinate(right) at ({ 2 * \a + \c * cos(\ang) + \delta }, { \c * sin(\ang) * 2.61 });
    
    \draw[dashed, line width=\linesWidth, white](left) -- (right);

    
	\def \refX {0.5 * \a + 0.5 * \c * cos(\ang)}
	\def \refY {0.5 * \c * sin(\ang)}
	
	\draw[->, line width=0.5pt, grey] ({\refX}, {\refY}) -- ({\refX + \a * 1}, {\refY}) node[left, xshift=3pt, yshift=-7pt] {\scriptsize $x$};
	\draw[->, line width=0.5pt, grey] ({\refX}, {\refY}) -- ({\refX}, {\refY + \c * sin(\ang) * 1}) node[left, , xshift=0pt, yshift=-1pt] {\scriptsize $y$};
    
	\draw[solid, line width=0.75pt, red](A1) -- (A2) -- (A3) -- (A4)--(A1);
	\foreach \vert in \verts {
		\filldraw [red] (\vert) circle (0.25pt); 
	}

    


    \foreach \ia in {0, 1} {
        \foreach \ic in {0, 1} {
            \def \centerX {\a / 2 + \a * \ia};
            \def \centerY {\c / 2 + \c * \ic};
            \foreach \i in {0, ..., 3} {
                \coordinate(this) at (
                    { \centerX + \RtoA * \a * cos(\ang * \i + \rot)}, 
                    { \centerY + \RtoA * \a * sin(\ang * \i + \rot)}
                );
                \filldraw [black, fill={\wCols[\i + 1]}, line width=0.25pt] (this) circle (\rw);
            }
        }
    }

    \foreach \ia in {0, 1} {
        \def \ic {-1};
        \def \centerX {\a / 2 + \a * \ia};
        \def \centerY {\c / 2 + \c * \ic};
        \foreach \i in {0, 1} {
            \coordinate(this) at (
                { \centerX + \RtoA * \a * cos(\ang * \i + \rot)}, 
                { \centerY + \RtoA * \a * sin(\ang * \i + \rot)}
            );
            \filldraw [black, fill={\wCols[\i + 1]}, line width=0.25pt] (this) circle (\rw);
        }
    }

    \foreach \ia in {0, 1} {
        \def \ic {2};
        \def \centerX {\a / 2 + \a * \ia};
        \def \centerY {\c / 2 + \c * \ic};
        \foreach \i in {2, 3} {
            \coordinate(this) at (
                { \centerX + \RtoA * \a * cos(\ang * \i + \rot)}, 
                { \centerY + \RtoA * \a * sin(\ang * \i + \rot)}
            );
            \filldraw [black, fill={\wCols[\i + 1]}, line width=0.25pt] (this) circle (\rw);
        }
    }

    \foreach \ic in {0, 1} {
        \def \ia {-1};
        \def \centerX {\a / 2 + \a * \ia};
        \def \centerY {\c / 2 + \c * \ic};
        \foreach \i in {0, 3} {
            \coordinate(this) at (
                { \centerX + \RtoA * \a * cos(\ang * \i + \rot)}, 
                { \centerY + \RtoA * \a * sin(\ang * \i + \rot)}
            );
            \filldraw [black, fill={\wCols[\i + 1]}, line width=0.25pt] (this) circle (\rw);
        }
    }

    \def \ia {-1}; \def \ic {-1};
    \def \centerX {\a / 2 + \a * \ia};
    \def \centerY {\c / 2 + \c * \ic};
    \foreach \i in {0} {
        \coordinate(this) at (
            { \centerX + \RtoA * \a * cos(\ang * \i + \rot)}, 
            { \centerY + \RtoA * \a * sin(\ang * \i + \rot)}
        );
        \filldraw [black, fill={\wCols[\i + 1]}, line width=0.25pt] (this) circle (\rw);
    }

    \def \ia {-1}; \def \ic {2};
    \def \centerX {\a / 2 + \a * \ia};
    \def \centerY {\c / 2 + \c * \ic};
    \foreach \i in {3} {
        \coordinate(this) at (
            { \centerX + \RtoA * \a * cos(\ang * \i + \rot)}, 
            { \centerY + \RtoA * \a * sin(\ang * \i + \rot)}
        );
        \filldraw [black, fill={\wCols[\i + 1]}, line width=0.25pt] (this) circle (\rw);
    }

    \foreach \ic in {0, 1} {
        \def \ia {2};
        \def \centerX {\a / 2 + \a * \ia};
        \def \centerY {\c / 2 + \c * \ic};
        \foreach \i in {1, 2} {
            \coordinate(this) at (
                { \centerX + \RtoA * \a * cos(\ang * \i + \rot)}, 
                { \centerY + \RtoA * \a * sin(\ang * \i + \rot)}
            );
            \filldraw [black, fill={\wCols[\i + 1]}, line width=0.25pt] (this) circle (\rw);
        }
    }

    \def \ia {2}; \def \ic {-1};
    \def \centerX {\a / 2 + \a * \ia};
    \def \centerY {\c / 2 + \c * \ic};
    \foreach \i in {1} {
        \coordinate(this) at (
            { \centerX + \RtoA * \a * cos(\ang * \i + \rot)}, 
            { \centerY + \RtoA * \a * sin(\ang * \i + \rot)}
        );
        \filldraw [black, fill={\wCols[\i + 1]}, line width=0.25pt] (this) circle (\rw);
    }

    \def \ia {2}; \def \ic {2};
    \def \centerX {\a / 2 + \a * \ia};
    \def \centerY {\c / 2 + \c * \ic};
    \foreach \i in {2} {
        \coordinate(this) at (
            { \centerX + \RtoA * \a * cos(\ang * \i + \rot)}, 
            { \centerY + \RtoA * \a * sin(\ang * \i + \rot)}
        );
        \filldraw [black, fill={\wCols[\i + 1]}, line width=0.25pt] (this) circle (\rw);
    }
    
	
	
	\def \circX {\a / 2 + 1 * \a + \RtoA * \a * cos(1 * \ang + \rot) + 1 * \c * cos(\ang)}
	\def \circY {\c / 2 + \RtoA * \a * sin(1 * \ang + \rot) + 1 * \c * sin(\ang)}
	\def \angArr {\ang + 45}

	\draw[->, black, line width=0.5pt]({\circX-3*\rw *cos(\angArr)}, {\circY-3*\rw * sin(\angArr)}) -- ({\circX-\rw*cos(\angArr)}, {\circY-\rw*sin(\angArr)}) node[midway,xshift=9pt, yshift=-4pt] {$2r_0$};
	\draw[->, black, line width=0.5pt]({\circX+3*\rw *cos(\angArr)}, {\circY+3*\rw * sin(\angArr)}) -- ({\circX+\rw*cos(\angArr)}, {\circY+\rw*sin(\angArr)});
	
    \def \deltaY {2.5 * \rw}
    
	\coordinate (aLeft) at (
        {\a}, {0}
    );
	\coordinate (aRight) at (
        {2 * \a}, {0}
    );
	\draw[<->, line width=0.5pt, black] (aLeft) -- (aRight) node[midway, below, sloped, xshift=0pt, yshift=1pt] {$a$};

    \coordinate (aLow) at (
        {0}, {\c}
    );
	\coordinate (aUp) at (
        {0}, {2 * \c}
    );
	\draw[<->, line width=0.5pt, black] (aLow) -- (aUp) node[midway, above, sloped, xshift=0pt, yshift=-1pt] {$a$};

    \def \deltaY {2.5 * \rw}
    
	\coordinate (aLeft) at (
        {\a / 2}, 
        {\c / 2}
    );
	\coordinate (aRight) at (
        {\a / 2 + \RtoA * \a * cos(\rot + 90)}, 
        {\c / 2 + \RtoA * \a * sin(\rot + 90)}
    );
	\draw[<->, line width=0.5pt, black] (aLeft) -- (aRight) node[midway, below, sloped, xshift=0pt, yshift=1pt] {$R$};

    \filldraw [black, fill=black, line width=0.25pt] (aLeft) circle (0.25pt);
    \filldraw [black, fill=black, line width=0.25pt] (aRight) circle (0.25pt);
	
\end{tikzpicture}
		}
	\end{minipage}
    \put(-117, 60){(d)}
    \put(-97, 60){II.4}
    \vspace{-0.4cm}
	\caption{
    Tunable geometries of wire media with (a-b) a hexagonal unit cell (general unit cell with $a=b$ and $\theta=\pi /3$, see Fig.~\ref{fig:gen_cell_0}) {containing six} 
    identical wires {with radius} 
    $r_0$; 
    (c-d) a square unit cell (general unit cell with $a=b$ and $\theta=\pi / 2$, see Fig.~\ref{fig:gen_cell_0}) {containing four} 
    identical wires {with radius} 
    $r_0$. 
	} \label{fig:geoms} 
\end{figure*}

Plasma-frequency tunability reported in recent studies \cite{kowitt2023tunable, balafendiev2022resonator, bae2023tunable} was limited to approximately $15$--$30\%$, where the tunability is defined as the ratio of the achievable frequency tuning range to its midpoint value.

Achieving substantially larger tunability remains a 
challenge. In this work, we address this problem by considering a breathing-deformation tuning mechanism for nested wire structures with $C_6$ and $C_4$ symmetries, illustrated in Fig.~\ref{fig:geoms}.
Moreover, we develop a theoretical framework for analyzing the dispersion properties of nested wire geometries containing $N\geq 2$ {different} wire {types} with radii $r_n$ located at positions $\boldsymbol{R}_n$ within the unit cell of an arbitrary two-dimensional periodic lattice, defined by lattice vectors $\boldsymbol{a}$ and $\boldsymbol{b}$ and the angle $\theta$ between them. By extending the local field approach \cite{BookTretyakov, belov2002dispersion} {to a general case of simple wire media}, we derive analytical expressions for the dispersion equation and discuss several limiting cases, including configurations in which closely spaced wires lead to numerical difficulties. For such cases, we introduce a \textit{cluster 
approximation} that significantly simplifies the analysis.

The developed model enables efficient estimation of the plasma frequency and its tunability for the nested wire geometries considered in this work.

\section{Theory}

To analyze the dispersion of wire media, we develop a general analytical framework based on the local field approach within the thin-wire approximation. In this section, we derive the dispersion equation for a two-dimensional periodic wire medium with an arbitrary lattice geometry and multiple wires within the unit cell.

From Maxwell's equations, the wave equation for the electric field $\boldsymbol{E}$ in SI units can be written as
\begin{equation}
    \nabla \times \nabla \times \boldsymbol{E}
    +
    \frac{1}{c^2} \frac{\partial^2}{\partial t^2} \boldsymbol{E}
    = 
    -\mu_0 \frac{\partial}{\partial t} \boldsymbol{J},
    \label{eq:wave_eq_0}
\end{equation}
where $\boldsymbol{J}$ is the current density distribution, $c$ is the speed of light, and $\mu_0$ is the permeability of free space.

We assume that all the wire radii $r_n$ are
significantly smaller than the lattice periods $a$ and $b$ ($r_n \ll a,b$). This allows us to approximate each wire as a $z$-oriented line current whose coordinates in the $xy$-plane coincide with the wire axis, instead of considering surface currents flowing at the wire interface. This approach is known as the \textit{line-current} or \textit{thin-wire approximation} \cite{belov2002dispersion}.

\begin{figure}[h!]
    \begin{minipage}{0.9\linewidth}
		\center{ 
			\input{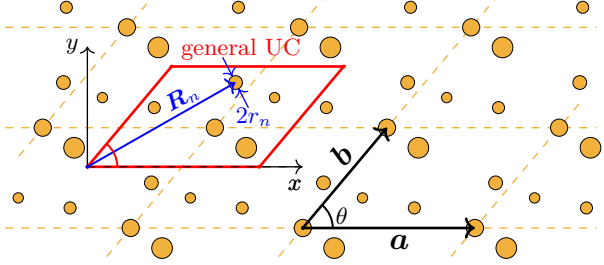}
		}
    \end{minipage}
	\caption{
    General case of a wire medium: {red solid line indicates} a \textit{general unit cell} 
    ({a parallelogram cell with} lattice vectors $\boldsymbol{a}=a \hat{\boldsymbol{a}}$ and $\boldsymbol{b}=b \hat{\boldsymbol{b}}$, where $\hat{\boldsymbol{a}}$ and $\hat{\boldsymbol{b}}$ are {the} unit vectors, $\angle (\hat{\boldsymbol{a}},\hat{\boldsymbol{b}})=\theta$) 
    {comprising multiple} wires ($N\geq 2$; 
    {this figure shows a medium with five different}  wires, but there can be more). 
    Each wire of radius $r_n$ {is} located {at the point} 
    $\boldsymbol{R}_n$.  
	} \label{fig:gen_cell_0} 
\end{figure}

We aim to solve the eigenvalue problem defined by Eq.~(\ref{eq:wave_eq_0}) for a general infinite periodic structure, illustrated in Fig.~\ref{fig:gen_cell_0}. The corresponding eigenvector is denoted by $\boldsymbol{q}$. Since the structure is periodic only in the $xy$-plane, all field quantities and currents share the same axial dependence $e^{-jq_z z}$. 

Assuming the time dependence $e^{j\omega t}$ ($\partial/\partial t = j\omega$) 
and taking into account the longitudinal dependence $e^{-jq_z z}$, Eq.~(\ref{eq:wave_eq_0}) is transformed into:
\begin{align}
    &
    \left(\nabla_t^2 + \frac{\partial^2}{\partial z^2}  + k^2 \right)
    \boldsymbol{E}
    (x,y)e^{-jq_zz}
    \label{eq:wave_eq} \\
    & \qquad\qquad 
    = 
    j\omega \mu_0 
    \left[
        \overline{\overline{\mathbb{I}}} + \frac{1}{k^2}\nabla \nabla
    \right]
    \boldsymbol{J}
    (x,y)e^{-jq_zz},
    \nonumber 
\end{align}
where $k=\omega/c$, $\nabla_t^2$ is the Laplacian in the $xy$-plane
and $\overline{\overline{\mathbb{I}}}$ is the unit dyad.
After applying the differential operators, the common longitudinal factor can be omitted. Thus, the analysis is performed in terms of the field and current amplitudes depending only on the transverse coordinates. The longitudinal dependence can be restored by multiplying the resulting fields by $e^{-jq_z z}$.

Equation (\ref{eq:wave_eq}) can be solved in cylindrical coordinates $(r,\phi,z)$ for a single line current $\boldsymbol{J}^\text{(w)} 
(\boldsymbol{R},\boldsymbol{R}^\prime)
=I\delta(\boldsymbol{R}-\boldsymbol{R}^\prime)\hat{\boldsymbol{z}}$ (where $\hat{\boldsymbol{z}}$ is a unit vector along the $z$ direction) with an amplitude $I$, using the scalar Green's function
\begin{equation}
    g(\boldsymbol{R}, \boldsymbol{R}^\prime)=\frac{j}{4}H_0^{(2)}\left(
        \sqrt{k^2-q_z^2} \;|\boldsymbol{R}-\boldsymbol{R}^\prime|
    \right).
    \label{eq:g_func}
\end{equation}
This function is an \textit{outgoing solution} satisfying the radiation condition for the scalar two-dimensional wave equation
\begin{equation}
    (\nabla_t^2 +\kappa^2) g(\boldsymbol{R},\boldsymbol{R}^\prime) = \delta(\boldsymbol{R}-\boldsymbol{R}^\prime),
    \label{eq:scalar_wave_eq}
\end{equation}
where $\kappa=\sqrt{k^2-q_z^2}$. The function $H_0^{(2)}$ in Eq.~(\ref{eq:g_func}) is the zero-order Hankel function of the second kind.

The dyadic Green's function describing the electric field generated by the current is
\begin{equation}
    \overline{\overline{G}}(\boldsymbol{R}, \boldsymbol{R}^\prime)=
    \left[
        \overline{\overline{\mathbb{I}}} + \frac{1}{k^2}
        \mathcal{D}\mathcal{D}
    \right]
    g(\boldsymbol{R},\boldsymbol{R}^\prime)
    ,
\end{equation}
where $\mathcal D=(\nabla_t-jq_z\hat{\boldsymbol{z}})$ is the differential operator accounting for the assumed longitudinal dependence.
Hence, the longitudinal component of the electric field generated by the current can be expressed as \cite{felsen1994radiation}:
\begin{align}
    E_z^\text{(w)}(\boldsymbol{R},\boldsymbol{R}^\prime) 
    =& j\omega \mu_0  
    \int_S
    G_{zz}
    (\boldsymbol{R}, \boldsymbol{R}^{\prime \prime} )
    J_{z}^\text{(w)} 
    (\boldsymbol{R}^{\prime \prime}, \boldsymbol{R}^\prime)
    dS^{\prime \prime}.
    \label{eq:ez_wire_full}
\end{align}
Then, the expression for the amplitude of the longitudinal component is:
\begin{align}
    E_z^\text{(w)}(\boldsymbol{R},\boldsymbol{R}^\prime) 
    =
    \frac{j \eta\kappa^2}{k} g(\boldsymbol{R}, \boldsymbol{R}^\prime)\;I,
    \label{eq:ez_wire}
\end{align}
where $\eta=\sqrt{\mu_0/\varepsilon_0}$ is the free-space wave impedance.


In the general case of the wire medium shown in Fig.~\ref{fig:gen_cell_0}, the 
coordinates of each wire with radius $r_n$ are given by $\boldsymbol{R}_n + \boldsymbol{R}^{(l,p)}$, where $\boldsymbol{R}^{(l,p)}=l\boldsymbol{a} + p \boldsymbol{b}$, and $l$ and $p$ are integers. For eigenstates with eigenvector $\boldsymbol{q}$ in an infinite periodic structure, the currents satisfy the plane-wave relation
\begin{equation}
    I_n^{(l,p)}=I_n e^{-j(\boldsymbol{q}, \boldsymbol{R}^{(l,p)})},
    \label{eq:currents_pw}
\end{equation}
where $I_n$ is the current amplitude in the wire located at $\boldsymbol{R}_n$ within the highlighted unit cell in Fig.~\ref{fig:gen_cell_0} ($l=0$, $p=0$), which we {call} 
the \textit{reference unit cell}.

A periodic Green's function for the infinite lattice can then be introduced as the lattice sum 
\begin{align}
    & g^\text{(lat)}(\boldsymbol{a},\boldsymbol{b}; \boldsymbol{R}, \boldsymbol{R}^\prime; k,\boldsymbol{q}) = 
    \nonumber \\
    &
    \sum\limits_{l=-\infty}^{+\infty}
    \sum\limits_{p=-\infty}^{+\infty}
    g \left(
        \boldsymbol{R}, \boldsymbol{R}^\prime + \boldsymbol{R}^{(l,p)}
    \right)
    e^{-j \left(\boldsymbol{q}, \boldsymbol{R}^{(l,p)}\right)}.
    \label{eq:green_x}
\end{align}

This function allows 
{calculating} the longitudinal electric field amplitude at a point $\boldsymbol{r} \neq \boldsymbol{R}^{(l,p)}$ in {the} $xy$-plane (the points $\boldsymbol{R}^{(l,p)}$ are singularity points) within the wire medium analogously to Eq.~(\ref{eq:ez_wire}):
\begin{align}
    E_z^\text{(wm)} (\boldsymbol{R})
    &=
    \frac{j\eta \kappa^2}{k}
    \sum\limits_{n=1}^N
    \sum\limits_{l,p}
    g(\boldsymbol{R}, \boldsymbol{R}_n+\boldsymbol{R}^{(l,p)}) \; I_n^{(l,p)}
    \nonumber \\
    &=
    \frac{j\eta \kappa^2}{k}
    \sum\limits_{n=1}^N
    g^\text{(lat)}(\boldsymbol{a},\boldsymbol{b}; \boldsymbol{R}, \boldsymbol{R}_n; k,\boldsymbol{q}) \; I_n 
    \nonumber \\
    & =
    \sum\limits_{n=1}^N
    D(a,b,\theta; \boldsymbol{R} - \boldsymbol{R}_n; k,\boldsymbol{q}) \; I_n,
    \label{eq:ez_wm_r}
\end{align}
where the \textit{interaction constant} $D$ is introduced to represent the field distribution. The open-form expressions for this constant are discussed in Appendix \ref{app:formulae}. To simplify the notation, the explicit dependence of the constant $D$ on the lattice parameters ($a$, $b$, and $\theta$) and on the wave vectors ($k$ and $\boldsymbol{q}$) will be omitted in the following expressions whenever no ambiguity arises.

The points {on} 
the surface of the $n^\text{th}$ wire in the reference unit cell are denoted by $\boldsymbol{R}_n^\text{(surf)}$, where $\boldsymbol{R}_n^\text{(surf)} - \boldsymbol{R}_n=\boldsymbol{r}_n$ and $|\boldsymbol{r}_n|=r_n$. The perfect electric conductor boundary condition for the $n^\text{th}$ wire requires the longitudinal electric field to vanish at its surface:
\begin{align}
    \sum\limits_{m\neq n}
    D(\boldsymbol{R}_n^\text{(surf)} - \boldsymbol{R}_m) \; I_m 
    +
    D(\boldsymbol{r}_n)
    \; I_n= 0.
    \label{eq:ez_wm_mth_wire}
\end{align}

Within the thin-wire approximation, the wire radius $r_n$ is assumed to be much smaller than the lattice periods, which also implies the inequality $r_n \ll |\boldsymbol{R}^{(l,p)}|$ for all pairs $(l,p) \neq (0,0)$. Hence, the argument $\boldsymbol{R}_n^\text{(surf)} - \boldsymbol{R}_m$ in the infinite sum of Eq.~(\ref{eq:ez_wm_mth_wire}) can be closely approximated by the inter-wire vector $\boldsymbol{\Delta}_{mn} = \boldsymbol{R}_n - \boldsymbol{R}_m$.

The last term on the left-hand side of Eq.~(\ref{eq:ez_wm_mth_wire}) can be written explicitly as
\begin{align}
    D(\boldsymbol{r}_n)\;
    I_n = &
    \nonumber \\
    \frac{j\eta\kappa^2}{k}
    \Bigg[ &
    \sum\limits_{(l,p)\neq (0,0)}
    g \left(
        \boldsymbol{R}_n, \boldsymbol{R}_n+\boldsymbol{R}^{(l,p)}
    \right)
    e^{-j \left(\boldsymbol{q}, \boldsymbol{R}^{(l,p)}\right)}
    \nonumber \\
    &+ 
    g \left(
        \boldsymbol{R}_n^\text{(surf)}, \boldsymbol{R}_n
    \right) 
    \Bigg]I_n.
    \label{eq:boundary_single_lat_2}
\end{align}

The sum over $(l,p)\neq (0,0)$ in Eq.~(\ref{eq:boundary_single_lat_2}) describes the contribution of all surrounding wires belonging to the same periodic sub-lattice to the longitudinal electric field amplitude {on} 
the surface of the considered $n^\text{th}$ wire in the reference unit cell. This sum is called the \textit{interaction constant} $C(a,b,\theta; k, \boldsymbol{q})$ \cite{belov2002dispersion}  (all arguments will be omitted in further expressions). In Appendix \ref{app:formulae}, two alternative methods to calculate this interaction constant are presented.

The second term of Eq.~(\ref{eq:boundary_single_lat_2}) can be rewritten in terms of 
\textit{effective susceptibility} \cite{belov2002dispersion}, where the inverse effective susceptibility $\alpha^{-1}$ of a wire with radius $r_n$ is defined as
\begin{align}
    \alpha^{-1}(r_n, k, q_z) &= 
    -\frac{j\eta\kappa^2}{k}
    g(
    \boldsymbol{R}_n^\text{(surf)}, \boldsymbol{R}_n
    )
    \nonumber \\
    & \approx \frac{\eta \kappa^2}{4k} 
	\left(
	1 - \frac{2j}{\pi} \left\{
	\log \frac{\kappa r_n}{2} + \gamma
	\right\}
	\right),
	\label{eq:eff_suscept}
\end{align}
where an asymptotic expansion of the Hankel function for small arguments was used, and $\gamma\approx 0.5772$ is the Euler-Mascheroni constant.

Consequently, within the line-current approximation, the perfect conductor boundary condition for the $n^\text{th}$ wire can be written compactly as
\begin{align}
\alpha_n^{-1}I_n=
C\; I_n
+ \sum_{m\neq n} D (\boldsymbol{\Delta}_{mn})\;I_m,
\label{eq:boundary_eq}
\end{align}
where $\alpha_n=\alpha(r_n)$ is the effective susceptibility of the $n^\text{th}$ wire.

Therefore, the system of boundary conditions for all $N$ wires {within} 
the unit cell (see Fig.~\ref{fig:gen_cell_0}) can be expressed in matrix form as
\begin{align}
     \left(
     \mathbb{A}^{-1} - \mathbb{C}
     \right) \cdot \mathbf{I} = 0,
    \label{eq:system_n}
\end{align}
where $\mathbf{I}$ is {a} 
vector of length $N$ containing the current amplitudes for each wire in the reference unit cell. 
The matrix $\mathbb{A}$ is a diagonal matrix of the effective wire susceptibilities,
\begin{equation}
    \mathbb{A} = \operatorname{diag} \left( \alpha_1,\dots, \alpha_N \right),
    \label{eq:def_a}
\end{equation}
while the interaction matrix $\mathbb{C}$ is a square matrix defined by elements
\begin{align}
    \mathbb{C}_{nm} = 
    \delta_{nm} C
    + 
    (1 - \delta_{nm})
    D
    (\boldsymbol{\Delta}_{mn}),
    \label{eq:def_c}
\end{align}
where $\delta_{nm}$ is the Kronecker delta.

For ease of reference, we introduce the notation $\mathbb{M} = \mathbb{A}^{-1} - \mathbb{C}$.
The dispersion equation for the wire structure shown in Fig.~\ref{fig:gen_cell_0} is obtained by setting the determinant of Eq.~(\ref{eq:system_n}) to zero, which yields the non-trivial solutions of the homogeneous system of boundary equations:
\begin{equation}
    \det \mathbb{M} = 0.
    \label{eq:det_N}
\end{equation}

When some wires are too close to each other{,} the 
convergenc{e} 
of the interaction constant $D$ {becomes problematic}. 
Hence, the \textit{cluster 
approximation} {allows obtaining} 
stable solutions of the eigen{problem} 
with an adequate precision.
The approximation 
{significantly reduces} the number of lattice sums required to construct the matrix $\mathbb{M}$ when multiple wires form closely spaced groups (or \textit{clusters}), as illustrated in Fig.~\ref{fig:gen_cell_clusters}, which {is} discussed in Appendix \ref{app:clusters}. 
Matrix (\ref{eq:system_clusters}) describes the general system of wires and demonstrates how it can be simplified in the presence of wire clusters. It directly coincides with {the} matrix $\mathbb{M}$ if each wire is treated as an individual, separate cluster.

It should be noted that the results previously reported in \cite{belov2002dispersion} and \cite{kowitt2023tunable} demonstrate specific reduction cases of this generalized analytical framework. Alternatively, a slightly different approach was discussed in \cite{nicorovici1995photonic} for a general wire lattice, though the case of multiple sub-lattices was not considered there. 

In the next sections, 
we apply our developed analytical model to four tunable geometries and compare the analytical results with full-wave numerical simulations performed in COMSOL \cite{comsol}.

\section{Breathing Deformation of Nested Wire Media}

The first two proposed geometries {with $C_6$ symmetry}, shown in Figs. \ref{fig:geoms}(a) and \ref{fig:geoms}(b), are based on a triangular (hexagonal) lattice with lattice vectors of equal length $a$ and an included angle of $\pi/3$. The primitive parallelogram unit cell is highlighted by solid red lines, {and} 
the equivalent hexagonal unit cell 
by 
dashed {black} line{s}. 
Structures {with this geometry} 
are also called \textit{honeycomb} \cite{wu2015scheme} structures.
Each hexagonal unit cell contains six identical wires of radius $r_0$ positioned at an equal distance $R$ from the cell center. The lattice vectors and wire coordinates for these geometries are summarized in Table \ref{tab:coordinates}.

\begin{table}[h!]
  \centering
  \begin{tabular}{|r||c | c|c|c|c|}
    \hline
    \rule{0pt}{9pt}
    \textbf{Geometry} & 
    $\boldsymbol{a}$ & $\boldsymbol{b}$ & 
    $\theta$ &
    $n$ & $\boldsymbol{R}_n / R$\\
    \hline
    \hline 
    \textbf{I.6} &  
    $ \left(
    \begin{matrix} 1\\0 \end{matrix}\right)a$
    & 
    $ \left(
    \begin{matrix} 1\\\sqrt{3} \end{matrix}\right)
    a/2$
    & 
    $\pi/3$ & 
    $1..6$ &
    $ \left(
    \begin{matrix} \cos \frac{n\pi}{3}\\ \sin \frac{n\pi}{3}  \end{matrix}\right)
    $
    \\
    \hline
    \textbf{II.6} &  
    $ \left(
    \begin{matrix} 1\\0 \end{matrix}\right)a$
    & 
    $ \left(
    \begin{matrix} 1\\\sqrt{3} \end{matrix}\right)
    a/2$
    & 
    $\pi/3$ & 
    $1..6$ &  
    $ \left(
    \begin{matrix} \cos \frac{(2n-1)\pi}{6} \\ \sin \frac{(2n-1)\pi}{6} \end{matrix}\right)
    $
    \\
    \hline \hline
    \textbf{I.4} &  
    $ \left(
    \begin{matrix} 1\\0 \end{matrix}\right)a$
    & 
    $ \left(
    \begin{matrix} 0\\ 1 \end{matrix}\right)
    a$
    & 
    $\pi/2$ & 
    $1..4$ &
    $ \left(
    \begin{matrix} \cos \frac{n\pi}{2}\\ \sin \frac{n\pi}{2}  \end{matrix}\right)
    $
    \\
    \hline
    \textbf{II.4} &  
    $ \left(
    \begin{matrix} 1\\0 \end{matrix}\right)a$
    & 
    $ \left(
    \begin{matrix} 0\\ 1 \end{matrix}\right)
    a$
    & 
    $\pi/2$ & 
    $1..4$ &  
    $ \left(
    \begin{matrix} \cos \frac{(2n-1)\pi}{4} \\ \sin \frac{(2n-1)\pi}{4} \end{matrix}\right)
    $
    \\
    \hline
  \end{tabular}
  \caption{
  Geometrical parameters for each structure shown in Fig.~\ref{fig:geoms} and coordinates of the wires within the unit cell, $\boldsymbol{R}_n$, assuming that the coordinate origin is located at the center of the unit cell, as illustrated in Fig.~\ref{fig:geoms}.
  }
  \label{tab:coordinates}
\end{table}

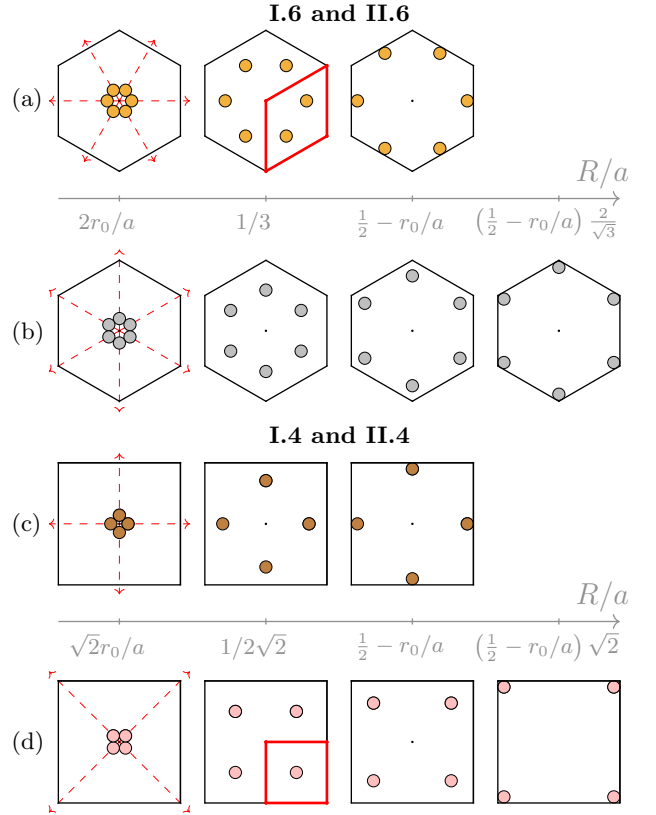
\begin{figure}[h!]
    \begin{minipage}{1.0\linewidth}
		\center{ 
			\definecolor{gold}{rgb}{0.95, 0.69, 0.24}
\definecolor{grey}{rgb}{0.57, 0.57, 0.57}

\begin{tikzpicture}[scale=0.95, transform shape]
    \def\numUC{3};  

    \def\cylcol{gold};
    
    
    \readlist \radiuses{0.1, 1/3, 0.45}; 
    \readlist*\labels{
    $2r_0/a$, $1/3$, $\frac{1}{2}-r_0/a$,
    $\left(\frac{1}{2} - r_0/a\right)\frac{2}{\sqrt{3}}$
    };
    
    \def\period{1.7};
    \def\rc{\period * 0.05};  
    \def\delta{0.2 * \period};  
    
    \readlist \sArr {0, 1, 2, 3};

    \def \len {1.15 * \period / 2}
    \foreach \i in {0, ..., 5} {
        \draw[dashed, ->, line width=0.05pt, red](0, 0) -- ({\len * cos(\i * 60)}, {\len * sin(\i * 60)});
    }
        
    \foreach \ind in {1, ..., 3} {

        \def \R {\radiuses[\ind]};
        \def \indThis {\sArr[\ind]};
        
        \foreach \i in {0, ..., 5} {
            \coordinate(this) at (
                {\indThis * (\period + \delta) + \period * \R * cos(\i * 60)}, 
                {\period * \R * sin(\i * 60)}
            );
            \coordinate(next) at (
                {\indThis * (\period + \delta) + \period * \R * cos((\i + 1) * 60)}, {\period * \R * sin((\i + 1) * 60)}
            );



            \filldraw [black, fill=\cylcol, line width=0.25pt] (this) circle (\rc);

        }

        \def\ucsz{\period / 2 / cos(30)};
        
        \foreach \i in {0, ..., 5} {
            \coordinate(this) at (
                {\indThis * (\period + \delta) + \ucsz * cos(30 + \i * 60)}, 
                {\ucsz * sin(30 + \i * 60)}
            );
            \coordinate(prev) at (
                {\indThis * (\period + \delta) + \ucsz * cos(30 + (\i - 1) * 60)},
                {\ucsz * sin(30 + (\i - 1) * 60)}
            );
            
            \draw[solid, line width=0.5pt, black](prev) -- (this);

            \filldraw [black] (this) circle (0.1pt); 
        }

        \coordinate (center) at ({(\ind - 1) * (\period + \delta)}, {0}); 
        \ifthenelse{\ind = 2}{
            \foreach \i in {5, 6} {
                \coordinate(this) at (
                    {\indThis * (\period + \delta) + \ucsz * cos(30 + \i * 60)}, 
                    {\ucsz * sin(30 + \i * 60)}
                );
                
                \coordinate(prev) at (
                    {\indThis * (\period + \delta) + \ucsz * cos(30 + (\i - 1) * 60)},
                    {\ucsz * sin(30 + (\i - 1) * 60)}
                );

                \ifthenelse{\i = 5}{
                    \draw[solid, line width=1.0pt, red](center) -- (prev);
                    \filldraw [red] (prev) circle (0.5pt); 
                }{}
                \ifthenelse{\i = 6}{
                    \draw[solid, line width=1.0pt, red](center) -- (this);
                }{}
                
                \draw[solid, line width=1.0pt, red](prev) -- (this);
                \filldraw [red] (this) circle (0.5pt); 
            }
        }{}
    }

    \foreach \ucInd in {0, ..., 2} {
        \filldraw [black] ({\ucInd * (\period + \delta)}, {0}) circle (0.25pt); 
    }

    \filldraw [red] ({1 * (\period + \delta)}, {0}) circle (0.5pt); 
    
    \def\downShift{0.8}  

    \def \margin {0.3}
    \draw[white] ({-\period * (\margin + 1/2)}, {-\downShift * \period}) -- ({\numUC * (\period + \delta) + \period * (\margin + 1/2)}, {-\downShift * \period});
    
    \draw[->, line width=0.5pt, grey] ({-\period / 2}, {-\downShift * \period}) -- ({\numUC * (\period + \delta) + \period / 2}, {-\downShift * \period}) node[left, xshift=7pt, yshift=10pt] {\large $R/a$};

    \def \dashSz {0.025 * \period};

    \foreach \ind in {1, ..., 4} {
    
        \def \xCord {\sArr[\ind] * (\period + \delta)}; 
        \draw[-, line width=0.5pt, grey] ({\xCord}, {-\downShift * \period + \dashSz}) -- ({\xCord}, {-\downShift * \period - \dashSz}) node[midway, xshift=-5pt, yshift=-10pt] {\labels[\ind]};

    }

\end{tikzpicture}
		}
        \put(-243, 53){(a)}
        \put(-146, 85){\textbf{I.6 and II.6}}
	\end{minipage}
    \vfill
	\begin{minipage}{1.00\linewidth}
		\center{
            \definecolor{gold}{rgb}{0.95, 0.69, 0.24}
\definecolor{grey}{rgb}{0.57, 0.57, 0.57}

\begin{tikzpicture}[scale=0.95, transform shape]
    \def\numUC{3};  

    \def\cylcol{lightgray};
    
    
    \def\period{1.7};
    \def\rc{\period * 0.05};  
    \def\delta{0.2 * \period};  

    \def \smallestR{\rc / \period * 2};
    \def \biggestR{0.5196};
    
    \readlist \radiuses{\smallestR, 1/3, 0.45, \biggestR}; 
    \readlist*\labels{$2r_0/a$, $1/3$, $\left(\frac{1}{2} - r_0/a\right)\frac{2}{\sqrt{3}}$};
    
    \readlist \sArr {0, 1, 2, 3};

    \def \len {1.15 * \period / 2 / cos(30)}
    \foreach \i in {0, ..., 5} {
        \draw[dashed, ->, line width=0.05pt, red](0, 0) -- ({\len * cos(\i * 60 + 30)}, {\len * sin(\i * 60 + 30)});
    }
    
    \foreach \ind in {1, ..., 4} {

        \def \R {\radiuses[\ind]};
        \def \indThis {\sArr[\ind]};
        
        \foreach \i in {0, ..., 5} {
            \coordinate(this) at (
                {\indThis * (\period + \delta) + \period * \R * cos(\i * 60 + 30)}, 
                {\period * \R * sin(\i * 60 + 30)}
            );
            \coordinate(next) at (
                {\indThis * (\period + \delta) + \period * \R * cos((\i + 1) * 60 + 30)}, {\period * \R * sin((\i + 1) * 60 + 30)}
            );



            \filldraw [black, fill=\cylcol, line width=0.25pt] (this) circle (\rc);
        }

        \def\ucsz{\period / 2 / cos(30)};
        
        \foreach \i in {0, ..., 5} {
            \coordinate(this) at (
                {\indThis * (\period + \delta) + \ucsz * cos(30 + \i * 60)}, 
                {\ucsz * sin(30 + \i * 60)}
            );
            \coordinate(prev) at (
                {\indThis * (\period + \delta) + \ucsz * cos(30 + (\i - 1) * 60)},
                {\ucsz * sin(30 + (\i - 1) * 60)}
            );
            
            \draw[solid, line width=0.5pt, black](prev) -- (this);

            \filldraw [black] (this) circle (0.1pt); 
        }
    }

    \foreach \ucInd in {0, ..., 3} {
        \filldraw [black] ({\ucInd * (\period + \delta)}, {0}) circle (0.25pt); 
    }
    
    \def\downShift{0.7}  

    \def \margin {0.3}
    \draw[white] ({-\period * (\margin + 1/2)}, {-\downShift * \period}) -- ({\numUC * (\period + \delta) + \period * (\margin + 1/2)}, {-\downShift * \period});
    
    \draw[->, line width=0.5pt, white] ({-\period / 2}, {-\downShift * \period}) -- ({\numUC * (\period + \delta) + \period / 2}, {-\downShift * \period});


    


\end{tikzpicture}
		}
        \put(-243, 32){(b)}
	\end{minipage}
    \vfill
    \vspace{0pt}
    \begin{minipage}{1.00\linewidth}
		\center{ 
			\begin{tikzpicture}[scale=0.95, transform shape]
    \def\numUC{3};  

    \def\cylcol{brown};
    
    
    \readlist \radiuses{0.0707, 0.35355, 0.45}; 
    \readlist*\labels{
    $\sqrt{2} r_0/a$, $1/2\sqrt{2}$, $\frac{1}{2}-r_0/a$,
    $\left(\frac{1}{2} - r_0/a\right)\sqrt{2}$
    };
    
    \def\period{1.7};
    \def\rc{\period * 0.05};  
    \def\delta{0.2 * \period};  
    
    \readlist \sArr {0, 1, 2, 3};
    \def \ang {90}
    \def \halfAng {\ang / 2}

    \def \len {1.15 * \period / 2}
    \foreach \i in {0, ..., 3} {
        \draw[dashed, ->, line width=0.05pt, red](0, 0) -- ({\len * cos(\i * \ang)}, {\len * sin(\i * \ang)});
    }
        
    \foreach \ind in {1, ..., 3} {

        \def \R {\radiuses[\ind]};
        \def \indThis {\sArr[\ind]};
        
        \foreach \i in {0, ..., 5} {
            \coordinate(this) at (
                {\indThis * (\period + \delta) + \period * \R * cos(\i * \ang)}, 
                {\period * \R * sin(\i * \ang)}
            );
            \coordinate(next) at (
                {\indThis * (\period + \delta) + \period * \R * cos((\i + 1) * \ang)}, {\period * \R * sin((\i + 1) * \ang)}
            );



            \filldraw [black, fill=\cylcol, line width=0.25pt] (this) circle (\rc);

        }

        \def\ucsz{\period / 2 / cos(\halfAng)};
        
        \foreach \i in {0, ..., 5} {
            \coordinate(this) at (
                {\indThis * (\period + \delta) + \ucsz * cos(\halfAng + \i * \ang)}, 
                {\ucsz * sin(\halfAng + \i * \ang)}
            );
            \coordinate(prev) at (
                {\indThis * (\period + \delta) + \ucsz * cos(\halfAng + (\i - 1) * \ang)},
                {\ucsz * sin(\halfAng + (\i - 1) * \ang)}
            );
            
            \draw[solid, line width=0.5pt, black](prev) -- (this);

            \filldraw [black] (this) circle (0.1pt); 
        }
    }

    \foreach \ucInd in {0, ..., 2} {
        \filldraw [black] ({\ucInd * (\period + \delta)}, {0}) circle (0.25pt); 
    }
    
    \def\downShift{0.8}  

    \def \margin {0.30}
    \draw[white] ({-\period * (\margin + 1/2)}, {-\downShift * \period}) -- ({\numUC * (\period + \delta) + \period * (\margin + 1/2)}, {-\downShift * \period});
    
    \draw[->, line width=0.5pt, grey] ({-\period / 2}, {-\downShift * \period}) -- ({\numUC * (\period + \delta) + \period / 2}, {-\downShift * \period}) node[left, xshift=7pt, yshift=10pt] {\large $R/a$};

    \def \dashSz {0.025 * \period};

    \foreach \ind in {1, ..., 4} {
    
        \def \xCord {\sArr[\ind] * (\period + \delta)}; 
        \draw[-, line width=0.5pt, grey] ({\xCord}, {-\downShift * \period + \dashSz}) -- ({\xCord}, {-\downShift * \period - \dashSz}) node[midway, xshift=-5pt, yshift=-10pt] {\labels[\ind]};

    }

\end{tikzpicture}
		}
        \put(-243, 52){(c)}
        \put(-146, 85){\textbf{I.4 and II.4}}
	\end{minipage}
    \vfill
	\begin{minipage}{1.00\linewidth}
		\center{
            \begin{tikzpicture}[scale=0.95, transform shape]
    \def\numUC{3};  

    \def\cylcol{pink};
    
    
    \readlist \radiuses{0.0707, 0.35355, 0.45, 0.6364};
    \readlist*\labels{
    $\sqrt{2}r_0/a$, $1/4$, $\frac{1}{2}-r_0/a$,
    $\left(\frac{1}{2} - r_0/a\right)\frac{2}{\sqrt{3}}$
    };
    
    \def\period{1.7};
    \def\rc{\period * 0.05};  
    \def\delta{0.2 * \period};  
    
    \readlist \sArr {0, 1, 2, 3};
    \def \ang {90}
    \def \halfAng {\ang / 2}

    \def \len {1.15 * \period / 2 / cos(\halfAng)}
    \foreach \i in {0, ..., 3} {
        \draw[dashed, ->, line width=0.05pt, red](0, 0) -- ({\len * cos(\i * \ang + \halfAng)}, {\len * sin(\i * \ang + \halfAng)});
    }
        
    \foreach \ind in {1, ..., 4} {

        \def \R {\radiuses[\ind]};
        \def \indThis {\sArr[\ind]};
        
        \foreach \i in {0, ..., 5} {
            \coordinate(this) at (
                {\indThis * (\period + \delta) + \period * \R * cos(\i * \ang + \halfAng)}, 
                {\period * \R * sin(\i * \ang + \halfAng)}
            );
            \coordinate(next) at (
                {\indThis * (\period + \delta) + \period * \R * cos((\i + 1) * \ang + \halfAng)}, {\period * \R * sin((\i + 1) * \ang + \halfAng)}
            );



            \filldraw [black, fill=\cylcol, line width=0.25pt] (this) circle (\rc);

        }

        \def\ucsz{\period / 2 / cos(\halfAng)};
        
        \foreach \i in {0, ..., 5} {
            \coordinate(this) at (
                {\indThis * (\period + \delta) + \ucsz * cos(\halfAng + \i * \ang)}, 
                {\ucsz * sin(\halfAng + \i * \ang)}
            );
            \coordinate(prev) at (
                {\indThis * (\period + \delta) + \ucsz * cos(\halfAng + (\i - 1) * \ang)},
                {\ucsz * sin(\halfAng + (\i - 1) * \ang)}
            );
            
            \draw[solid, line width=0.5pt, black](prev) -- (this);

            \filldraw [black] (this) circle (0.1pt); 
        }
    }

    \foreach \ucInd in {1, ..., 3} {
        \filldraw [black] ({\sArr[\ucInd] * (\period + \delta)}, {0}) circle (0.25pt); 
    }

    \coordinate(ul) at ({\period + \delta}, {0});
    \coordinate(ur) at ({\period + \delta + \period / 2}, {0});
    \coordinate(lr) at ({\period + \delta + \period / 2}, { -\period / 2});
    \coordinate(ll) at ({\period + \delta}, {-\period / 2});

    \readlist* \smallUC{ul, ur, lr, ll, ul};

    \foreach \ind in {1, ..., 4} {
        \draw[solid, line width=1.0pt, red](\smallUC[\ind]) -- (\smallUC[\ind + 1]);
        \filldraw [red] (\smallUC[\ind]) circle (0.5pt);
    } 
    
    
    \def\downShift{0.6}  

    \def \margin {0.30}
    \draw[white] ({-\period * (\margin + 1/2)}, {-\downShift * \period}) -- ({\numUC * (\period + \delta) + \period * (\margin + 1/2)}, {-\downShift * \period});
    


    


\end{tikzpicture}
		}
        \put(-243, 25){(d)}
	\end{minipage}
	\caption{
    Breathing deformation of the wire structures from Fig.~\ref{fig:geoms}, where the wires move toward (a, c) the 
    {edges} and (b, d) the vertices of the unit cell with an increase in the deformation parameter $R/a$.
	} \label{fig:deforms} 
\end{figure}

The remaining two structures exhibit $C_4$ symmetry, as shown in Figs. \ref{fig:geoms}(c) and \ref{fig:geoms}(d). These geometries consist of four identical wires with radius $r_0$ arranged within a square unit cell of size $a\times a$. Similar to the previous case, the wire positions are determined by the distance $R$ between the wire axes and the center of the unit cell. The corresponding wire coordinates are also provided in Table \ref{tab:coordinates}.


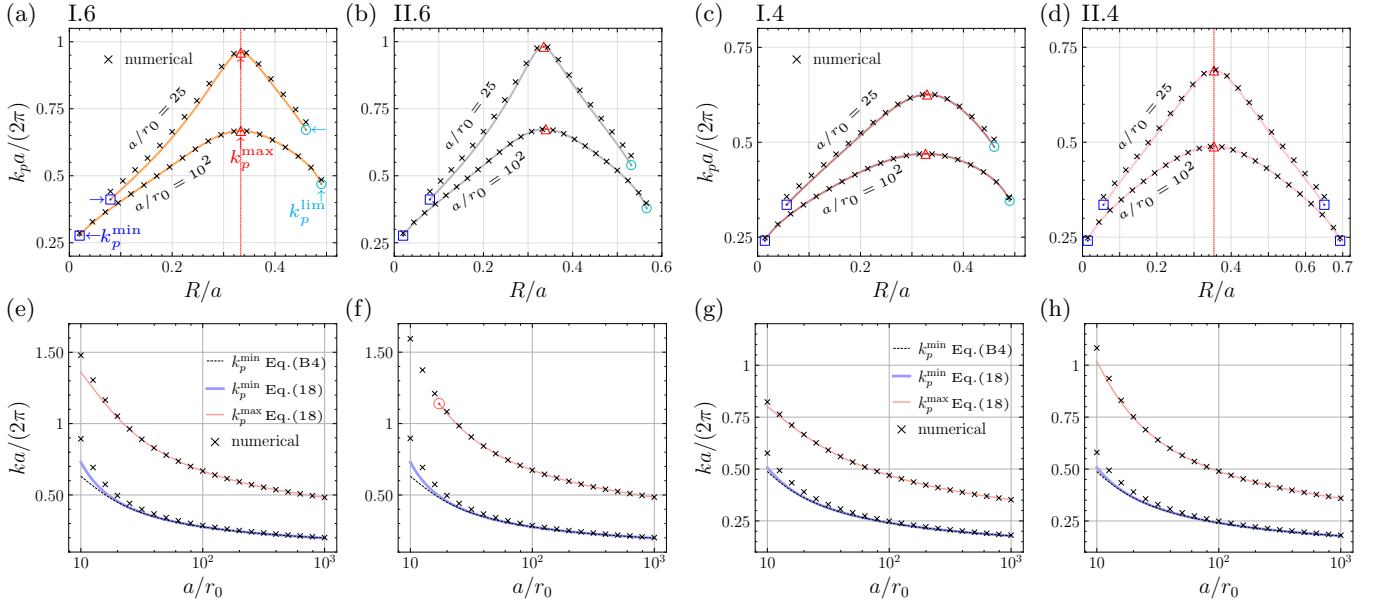
\begin{figure*}[t]
    \begin{minipage}{0.25\linewidth}
		\center{ 
			\includegraphics[width=1.00\textwidth]{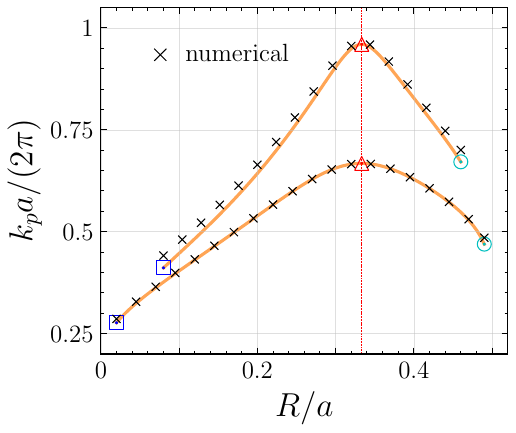}
		}
	\end{minipage}
    \put(-126.5, 62){(a)}
    \put(-103, 62){I.6}
    \put(-85, 13){
        \rotatebox{47}{\tiny $a/r_0=25$}
    }
    \put(-83, -10){
        \rotatebox{30}{\tiny $a/r_0=10^2$}
    }
    \put(-97, -20.2){\tiny \textcolor{blue}{$\leftarrow$}}
    \put(-95.5, -6.4){\tiny \textcolor{blue}{$\rightarrow$}}
    \put(-91, -21.2){\textcolor{blue}{\scriptsize $k_p^\text{min}$}}
    \put(-39.85, 15){\tiny \textcolor{red}{$\uparrow$}}
    \put(-39.85, 44.5){\tiny \textcolor{red}{$\uparrow$}}
    \put(-42.3, 8.5){\textcolor{red}{\scriptsize $k_p^\text{max}$}}
    \put(-9.5, -5){\tiny \textcolor{cyan}{$\uparrow$}}
    \put(-12, 19.6){\tiny \textcolor{cyan}{$\leftarrow$}}
    \put(-20, -12){\textcolor{cyan}{\scriptsize $k_p^\text{lim}$}}
    \hfill
	\begin{minipage}{0.237\linewidth}
		\center{
            \includegraphics[width=1.00\textwidth]{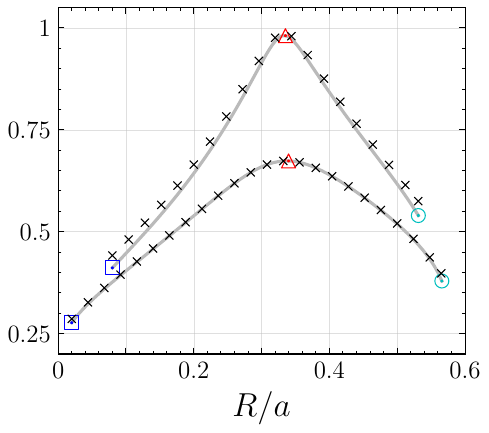}
		}
	\end{minipage}
    \put(-125, 62){(b)}
    \put(-106.7, 62){II.6}
    \put(-92, 13){
        \rotatebox{54}{\tiny $a/r_0=25$}
    }
    \put(-90, -10){
        \rotatebox{34}{\tiny $a/r_0=10^2$}
    }
    \hfill
	\begin{minipage}{0.25\linewidth}
		\center{
            \includegraphics[width=1.00\textwidth]{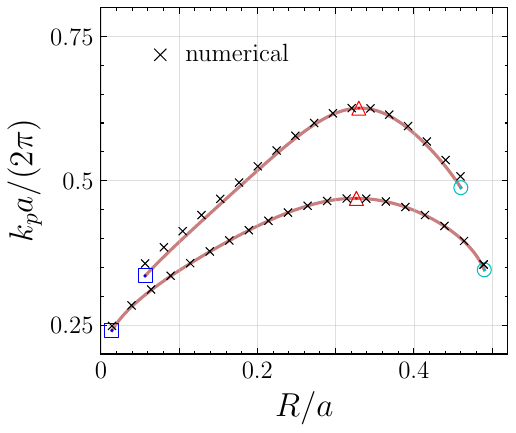}
		}
	\end{minipage}
    \put(-126.5, 62){(c)}
    \put(-103, 62){I.4}
    \put(-85, 13){
        \rotatebox{43}{\tiny $a/r_0=25$}
    }
    \put(-83, -11){
        \rotatebox{25}{\tiny $a/r_0=10^2$}
    }
    \hfill
	\begin{minipage}{0.2315\linewidth}
		\center{
            \includegraphics[width=1.00\textwidth]{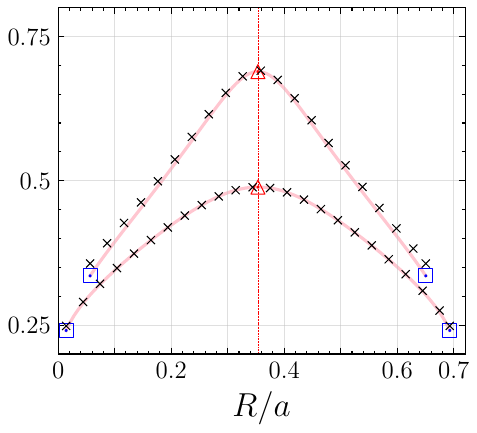}
		}
	\end{minipage}
    \put(-120, 62){(d)}
    \put(-104, 62){II.4}
    \put(-95, 13){
        \rotatebox{51}{\tiny $a/r_0=25$}
    }
    \put(-95, -13){
        \rotatebox{32}{\tiny $a/r_0=10^2$}
    }
    \vfill
    \vspace{-15pt}
    \begin{minipage}{0.25\linewidth}
		\center{ 
			\includegraphics[width=1.00\textwidth]{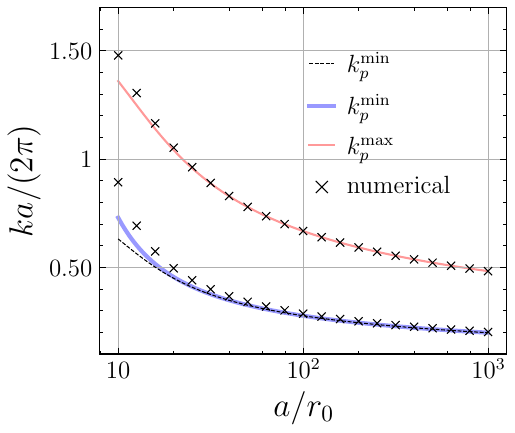}
		}
	\end{minipage}
    \put(-126.5, 62){(e)}
    \put(-29, 43){\tiny Eq.(\ref{eq:det_cluster})}
    \put(-29, 32.7){\tiny Eq.(\ref{eq:det_N})}
    \put(-29, 23){\tiny Eq.(\ref{eq:det_N})}
    \hfill
	\begin{minipage}{0.23\linewidth}
		\center{
            \includegraphics[width=1.00\textwidth]{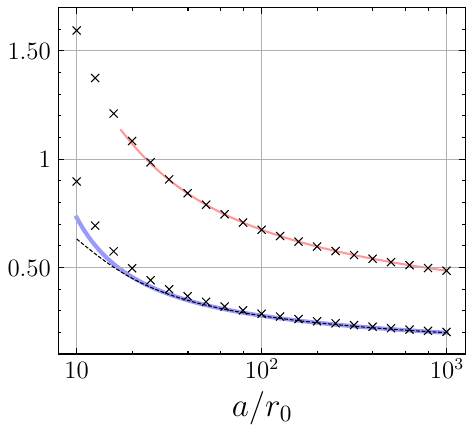}
		}
	\end{minipage}
    \put(-123, 62){(f)}
    \put(-90, 27.8){\tiny \textcolor{red}{$\odot$}}
    \hfill
	\begin{minipage}{0.25\linewidth}
		\center{
            \includegraphics[width=1.00\textwidth]{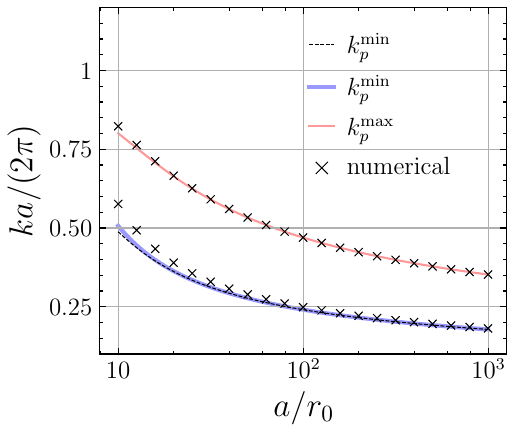}
		}
	\end{minipage}
    \put(-126, 62){(g)}
    \put(-29, 47.7){\tiny Eq.(\ref{eq:det_cluster})}
    \put(-29, 37.3){\tiny Eq.(\ref{eq:det_N})}
    \put(-29, 27.5){\tiny Eq.(\ref{eq:det_N})}
    \hfill
	\begin{minipage}{0.23\linewidth}
		\center{
            \includegraphics[width=1.00\textwidth]{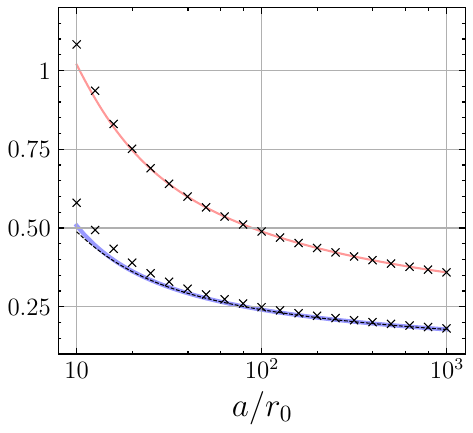}
		}
	\end{minipage}
    \put(-120, 62){(h)}
    \vspace{-0.5cm}
	\caption{
    (a-d) Plasma frequency tuning via breathing deformation of the structures from Fig.~\ref{fig:geoms}. Cross markers show numerically obtained plasma frequencies. Solid lines are analytical curves for the corresponding {wire systems.} 
    Upper curves in all subplots are plotted for $a/r_0=25$, {and lower ones} 
    for $a / r_0 = 100$.
    (e-h) Dependencies of $k_p^\text{max}$ and $k_p^\text{min}$ on wire radii. Cross markers were obtained numerically, while the solid and dashed lines were calculated analytically{,} with the corresponding equations designated in the legend.
	} \label{fig:freq_tuning} 
\end{figure*}

The plasma frequency for all {the} proposed geometries can be tuned by simultaneously changing the distance $R$ between each wire and the center of the unit cell, as illustrated in Fig.~\ref{fig:deforms}. This method is commonly referred to as \textit{breathing deformation} and is a well-established concept in topological photonics \cite{wu2015scheme, benalcazar2019quantization}.

For convenience, we label each structure using a combination of Roman (I or II) and Arabic (6 or 4) numerals. The Roman numeral indicates the direction of wire movement as the parameter $R$ increases: 
{``}I" means the wires move toward the unit cell 
{edges} [Figs. \ref{fig:deforms}(a) and \ref{fig:deforms}(c)], while 
{``}II" means they move toward the vertices [Figs. \ref{fig:deforms}(b) and \ref{fig:deforms}(d)]. The Arabic numeral represents the rotational symmetry of the structure{:} 
$6$ or $4$ for the $C_6$ and $C_4$-symmetric structures, respectively. 
{In the following,} we will refer to the structures according to this classification.

The geometric parameter $R$ varies within a specific range for each structure. For all geometries, 
$R$ {is minimal} 
when the wires are closely grouped near the unit cell center{, with} 
each wire {touching} 
its two nearest neighbors. For the $C_6$ and $C_4$ symmetric structures, this minimum value equals $2r_0$ and $\sqrt{2}r_0$, respectively. 
{$R$ reaches maximum} when the wires touch the unit cell boundaries and meet wires from adjacent cells (see Fig.~\ref{fig:geoms_lim}). The limit values of $R$ for all considered geometries are summarized in Table \ref{tab:r_limits} and visualized in Fig.~\ref{fig:deforms}.

\begin{table}[h!]
  \centering
  \begin{tabular}{|r||c|c|}
    \hline
    \rule{0pt}{9pt}
    \textbf{Geometry} & minimal $R$ & 
    maximal $R$ \\
    \hline
    \hline \rule{0pt}{12pt}
    I.6 & $2r_0$ & $a/2 - r_0$ \\
    \rule{0pt}{12pt}
    II.6 & $2r_0$ & $a/\sqrt{3} - 2r_0/\sqrt{3}$ \\
    \rule{0pt}{12pt}
    I.4 & $\sqrt{2}r_0$ & $a/2 - r_0$ \\
    \rule{0pt}{12pt}
    II.4 & $\sqrt{2}r_0$ & $a/\sqrt{2} - \sqrt{2}r_0$ \\
    \hline
  \end{tabular}
  \caption{
  Ranges for the geometric parameter $R$ for each geometry from Fig.~\ref{fig:geoms}.
  }
  \label{tab:r_limits}
\end{table}

There are two special cases, shown in Fig.~\ref{fig:deforms}(a) and (d), that occur during the breathing deformation of geometries I.6 and II.4. When $R$ equals $R^{(\text{I.6})}=a/3$ and $R^{(\text{II.4})}=a/2\sqrt{2}$, respectively, the periodicity of the structures changes. Specifically, for geometry I.6, the unit cell becomes three times smaller, while for geometry II.4, the period is reduced by half. These cases are highlighted in Figs. \ref{fig:deforms}(a) and (d) with solid red polygons indicating the actual unit cells for 
{these} specific $R/a$ values. 
For geometry II.4, the value $R^{(\text{II.4})}$ also represents a symmetry point for the breathing deformation; any configuration with $R_1\leq R^{(\text{II.4})}$ is identical to a configuration where $R_2=2R^{(\text{II.4})} - R_1$.

\section{Tuning of the plasma frequency}

{The described} breathing deformation enables effective tuning of the plasma frequency, as demonstrated in Figs. \ref{fig:freq_tuning}(a)--\ref{fig:freq_tuning}(d). 

The cross markers show the numerically calculated values {of} 
the lowest eigenfrequency at the $\Gamma$-point versus the geometric parameter $R$. For visual clarity, fewer numerical data points are plotted than were actually calculated, {with} 
a simulation step of $\Delta(R/a) = 10^{-3}$. Since the plasma frequency also depends strongly on the wire radius \cite{pendry1996extremely, belov2002dispersion, maslovski2009nonlocal, kumar2012novel}, two distinct sets of data points are plotted for each geometry: the lower curve corresponds to $a/r_0=100${,} and the upper curve 
to $a/r_0=25$. The solid lines represent the analytical results.

All configurations exhibit a similar general trend. When the wires are tightly grouped and touch each other near the center of the unit cell, the system reaches its minimum plasma wavenumber $k_p^\text{min}$ [
blue square markers in Figs. \ref{fig:freq_tuning}(a)--\ref{fig:freq_tuning}(d)]. As the deformation parameter $R$ increases, the plasma wavenumber rises up to a maximum value $k_p^\text{max}$ [red triangular markers in Figs. \ref{fig:freq_tuning}(a)--\ref{fig:freq_tuning}(d)]. Beyond this peak, the plasma frequency decreases down to a limiting wavenumber $k_p^\text{lim} > k_p^\text{min}$ [cyan circle markers in Figs. \ref{fig:freq_tuning}(a)--\ref{fig:freq_tuning}(d)] 
when the wires meet the boundaries of adjacent cells, either at the unit cell 
{edges} for geometries I.6 and I.4 or at the vertices for geometries II.6 and II.4 (see Fig.~\ref{fig:geoms_lim}).

To quantify the plasma frequency tuning range achieved via the breathing deformation, we define the \textit{tunability value} $\xi$ as:
\begin{equation}
    \xi = 2\;\frac{k_p^\text{max} - k_p^\text{min}}{k_p^\text{max} + k_p^\text{min}}.
    \label{eq:tunability}
\end{equation}

While the minimum plasma wavenumber {is always achieved at} 
the smallest possible value of $R/a$ regardless of the wire thickness, the exact {$R/a$ value providing the maximum plasma wavenumber} 
depends on the specific geometry.

\begin{table}[h!]
  \centering
  \begin{tabular}{| c|| c|c|c || c|c|c ||}
    \multicolumn{1}{c}{}
    & \multicolumn{6}{c}{ $k_p^\text{min}a/(2\pi)$ } 
    \\
    \hline \rule{0pt}{9pt}
    \multirow{2}{*}{$a/r_0$} 
    & \multicolumn{3}{c||}{ \textbf{I.6 and II.6} } 
    & \multicolumn{3}{c||}{ \textbf{I.4 and II.4} }
    \\
    \cline{2-7}
    \rule{0pt}{9pt}
    & num. & 
    Eq. (\ref{eq:det_cluster}) &
    $\Delta k_p (\%)$
    & num. & Eq. (\ref{eq:det_cluster}) &
    $\Delta k_p (\%)$
    \\ 
    \hline
    \hline
    20 & 
    0.4951 & 0.4492 & $9.27$ &    
    0.3894 & 0.3614 & $7.19$
    \\
    25 & 
    0.4415 & 0.4090  & $7.36$ &    
    0.3567 & 0.3348 & $6.14$  
    \\
    50 & 
    0.3411 & 0.3256 & $4.54$ &    
    0.2892 & 0.2771 & $4.18$  
    \\
    100 & 
    0.2861 & 0.2767 & $3.29$ &    
    0.2487 & 0.2407 & $3.22$  
    \\
    200 & 
    0.2508 & 0.2443 & $2.59$ &    
    0.2212 & 0.2155 & $2.58$  
    \\
    1000 & 
    0.2018 & 0.1984 & $1.68$ &    
    0.1812 & 0.1780 & $1.77$ 
    \\
    \hline
  \end{tabular}
  \caption{
  {Minimum plasma wavenumbers} $k_p^\text{min}$ 
  calculated numerically in COMSOL and {analytically with} 
  Eq.~(\ref{eq:det_cluster}) for different $a/r_0$ values.
  }
  \label{tab:estimations_min}
\end{table}

Some representative 
$k_p^\text{min}$ {values} corresponding to the leftmost points in Figs.~\ref{fig:freq_tuning}(a)--\ref{fig:freq_tuning}(d) are provided in Table \ref{tab:estimations_min}. As the ratio $a/r_0$ increases (i.e., for thinner wires), the relative error of the analytical estimation decreases, reaching approximately $\sim 2\%$ for $a/r_0=1000$.

\begin{figure}[h!]
    \begin{minipage}{0.52\linewidth}
		\center{ 
			\includegraphics[width=1.00\textwidth]{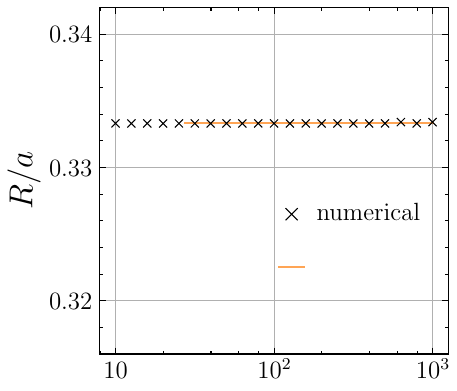}
		}
	\end{minipage}
    \put(-126, 64){(a)}
    \put(-99.5, 64){I.6}
    \put(-39, -14){\tiny Eq.(\ref{eq:det_N})}
    \put(-78.2, 26.2){\tiny \textcolor{red}{$\odot$}}
    \hfill
	\begin{minipage}{0.4683  \linewidth}
		\center{
            \includegraphics[width=1.00\textwidth]{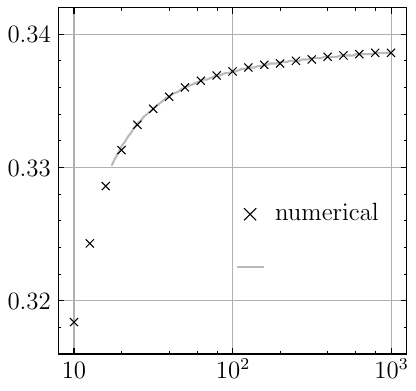}
		}
	\end{minipage}
    \put(-115, 64){(b)}
    \put(-99.5, 64){II.6}
    \put(-39, -14){\tiny Eq.(\ref{eq:det_N})}
    \put(-86.9, 14.5){\tiny \textcolor{red}{$\odot$}}
    \vspace{-0.3cm}
    \begin{minipage}{0.52\linewidth}
		\center{ 
			\includegraphics[width=1.00\textwidth]{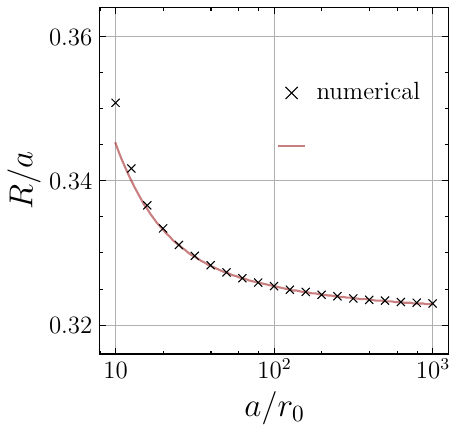}
		}
	\end{minipage}
    \put(-126, 69){(c)}
    \put(-99.5, 69){I.4}
    \put(-39, 25.5){\tiny Eq.(\ref{eq:det_N})}
    \hfill
	\begin{minipage}{0.4683  \linewidth}
		\center{
            \includegraphics[width=1.00\textwidth]{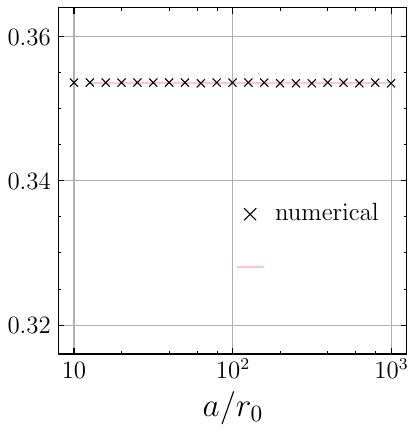}
		}
	\end{minipage}
    \put(-115, 69){(d)}
    \put(-99.5, 69){II.4}
    \put(-39, -8.2){\tiny Eq.(\ref{eq:det_N})}
    \put(-93, 43.3){\tiny \textcolor{red}{$\odot$}}
    \vspace{-0.6cm}
	\caption{
    Dependence of the $R/a$ value {corresponding to} 
    $k_p^\text{max}$ 
    on the wire radius $r_0$ for (a-b) honeycomb breathing structures (Fig.~\ref{fig:geoms}(a-b)) and (c-d) square breathing structures (Fig.~\ref{fig:geoms}(c-d)). Solid lines represent the analytically verified maximum positions calculated with an $R/a$ axis step size of $10^{-4}$. The cross markers were numerically obtained in COMSOL with the same step along $R/a$ axis. 
	} \label{fig:r_for_kmax}
\end{figure}
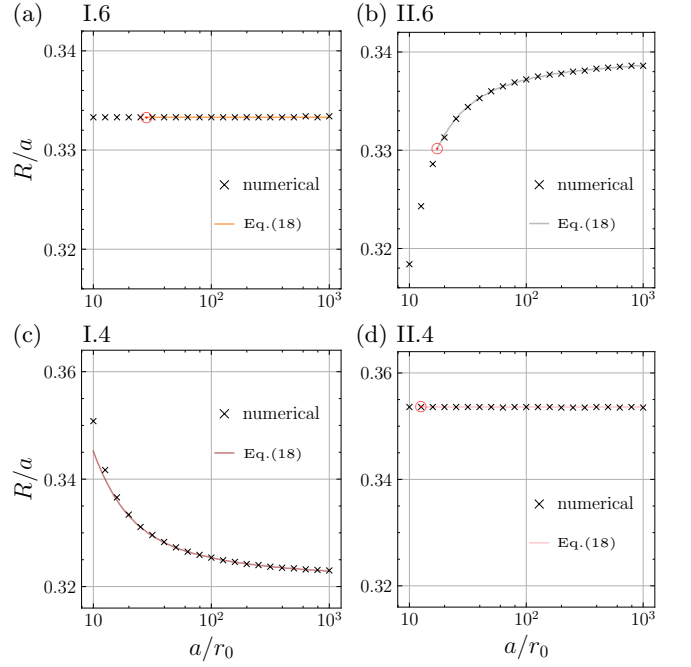

We performed numerical simulations to identify the position of the maximum plasma frequency for various wire radii. The results for each geometry are shown as cross markers in Fig.~\ref{fig:r_for_kmax}, where the numerical data step along the $R/a$ axis is $\Delta(R/a) = 10^{-4}$. 
The solid lines represent the 
positions of these extrema {analytically calculated} using Eq.~(\ref{eq:det_N}) with matrix $\mathbb{M}$ composed for all wires in the geometries (without simplifying), which show good agreement with the numerical data. 
For large wire radii, the validity of the analytical model is {limited for} 
certain geometries because the first root at the $\Gamma$-point of Eq.~(\ref{eq:det_N}) approaches the Bragg resonances. Consequently, the analytical curves are plotted only within the range of $R/a$ where the model remains applicable for each geometry.

\begin{figure*}[t]
    \begin{minipage}{0.26\linewidth}
		\center{ 
			\includegraphics[width=1.00\textwidth]{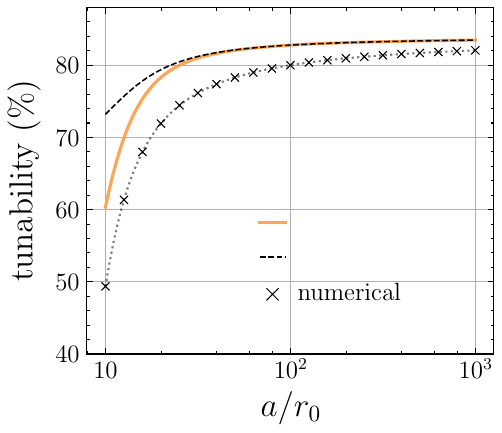}
		}
	\end{minipage}
    \put(-132.5, 66){(a)}
    \put(-110, 66){I.6}
    \put(-54.5, 4){\tiny Eq.(\ref{eq:det_N})}
    \put(-54.5, -5){\tiny Eqs.(\ref{eq:det_cluster},\ref{eq:det_N})}
    \hfill
	\begin{minipage}{0.235\linewidth}
		\center{
            \includegraphics[width=1.00\textwidth]{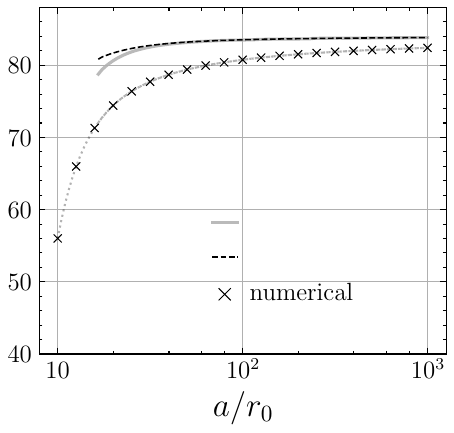}
		}
	\end{minipage}
    \put(-123, 66){(b)}
    \put(-110, 66){II.6}
    \put(-54.5, 4){\tiny Eq.(\ref{eq:det_N})}
    \put(-54.5, -5){\tiny Eqs.(\ref{eq:det_cluster},\ref{eq:det_N})}
    \put(-96.7, 43.7){\tiny \textcolor{red}{$\odot$}}
    \put(-96.7, 47.7){\tiny \textcolor{red}{$\odot$}}
    \put(-54.5, 4){\tiny Eq.(\ref{eq:det_N})}
    \put(-54.5, -5){\tiny Eqs.(\ref{eq:det_cluster},\ref{eq:det_N})}
    \hfill
	\begin{minipage}{0.26\linewidth}
		\center{
            \includegraphics[width=1.00\textwidth]{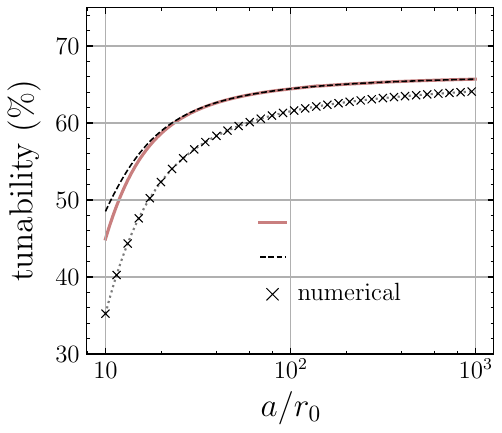}
		}
	\end{minipage}
    \put(-132, 66){(c)}
    \put(-110, 66){I.4}
    \put(-54.5, 4){\tiny Eq.(\ref{eq:det_N})}
    \put(-54.5, -5){\tiny Eqs.(\ref{eq:det_cluster},\ref{eq:det_N})}
    \hfill
	\begin{minipage}{0.235\linewidth}
		\center{
            \includegraphics[width=1.00\textwidth]{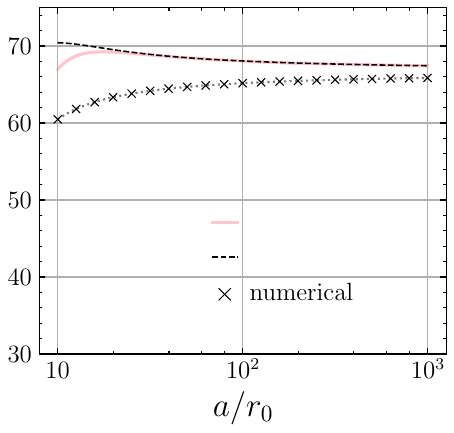}
		}
	\end{minipage}
    \put(-123, 66){(d)}
    \put(-110, 66){II.4}
    \put(-54.5, 4){\tiny Eq.(\ref{eq:det_N})}
    \put(-54.5, -5){\tiny Eqs.(\ref{eq:det_cluster},\ref{eq:det_N})}
    \vspace{-0.5cm}
	\caption{
    Dependence of the {estimated} maximal tunability 
    (achiev{ed} 
    via the breathing deformation of the structures from Fig.~\ref{fig:geoms}) on the wire radius $r_0$.
	} \label{fig:tunability} 
\end{figure*}
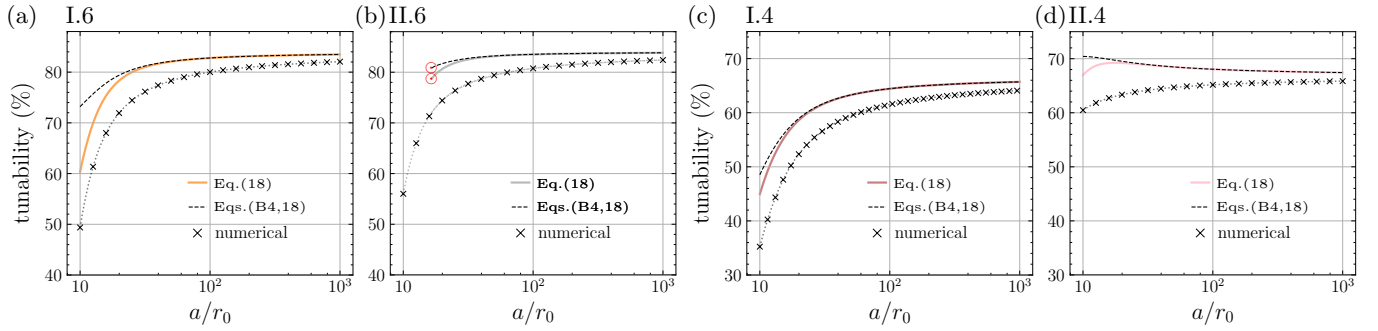

For geometries I.6 and II.4, the 
$R/a$ {value} at which $k_p^\text{max}$ is achieved is entirely determined  by the symmetry of the  lattice and does not depend on the wire radius [see Figs.~\ref{fig:r_for_kmax}(a) and \ref{fig:r_for_kmax}(d)]. These values are equal to $1/3$ and $1/(2\sqrt{2})$, respectively, which correspond to configurations with reduced structural periodicity [see Figs.~\ref{fig:deforms}(a) and \ref{fig:deforms}(d)]. Consequently, the maximum plasma frequency for these two geometries can be estimated analytically by considering simplified, equivalent smaller unit cells containing only two wires or one wire, respectively. This simplification significantly reduces the computational effort required to construct the matrix $\mathbb{M}$ by reducing its size from $6\times6$ to $2\times2$ and from $4\times4$ to $1\times1$, respectively.

For the other two geometries, II.6 and I.4, the position of the maximum plasma frequency during the breathing deformation depends on the wire radius{,} as shown in Figs.~\ref{fig:r_for_kmax}(b) and \ref{fig:r_for_kmax}(c).

Based on these calculated extrema locations, we evaluated the minimum and maximum achievable plasma wavenumbers, $k_p^\text{min}$ and $k_p^\text{max}$, for each geometry. In Figs.~\ref{fig:freq_tuning}(e)--\ref{fig:freq_tuning}(h), the cross markers represent the numerical results obtained via COMSOL, {and} 
the solid blue and red lines{,} 
the analytical values calculated by solving the complete system of equation{s} (\ref{eq:boundary_eq}) ({we calculated} 
all interaction constants $D$ entering the matrix $\mathbb{M}$ 
by selecting the \textit{appropriate} analytical expression, including the approximation valid for small $\boldsymbol{r}$, see Appendices \ref{app:formulae} and \ref{app:clusters}). {These numerical and analytical results are in} 
good agreement.

The dashed lines in each plot {are} 
the analytical results for $k_p^\text{min}$ obtained by treating the grouped wires near the unit cell center ($\boldsymbol{\Delta}_{mn} \sim r_0$) as a single closely spaced cluster. 
These dashed curves are calculated using the approximation (\ref{eq:cluster_matrix_gamma}) for the matrix $\mathbb{M}$, 
derived under the assumption of small wire radii within the single-cluster approximation detailed in Appendix \ref{app:clusters}. For larger wire radii ($a/r_0 < 20$), this approximation deviates more significantly from the numerical results. In this regime, neither the thin-wire approximation ($r_0 \ll a,b$) nor the cluster approximation ($\boldsymbol{\Delta}_{mn} \gg r_0$) remains strictly valid.

The tunability value (\ref{eq:tunability}) was then evaluated for each geometry as a function of the wire radius using the data from Figs. \ref{fig:freq_tuning}(e)--\ref{fig:freq_tuning}(h). Figure \ref{fig:tunability} shows the numerical results (cross markers with a dotted line; showing a subset of the $100$ calculated points along the $a/r_0$ axis) along with the analytically estimated tunability percentages (thick solid lines and dashed lines). 
$k_p^\text{min}$ and $k_p^\text{max}$ 
calculated by solving Eq.~(\ref{eq:det_N}) for both configurations {are shown with solid lines, and $k_p^\text{min}$ calculated by} 
Eq.~(\ref{eq:det_cluster}){, with dashed lines.}

For both $C_6$ geometries, the tunability increases as the wire radius decreases, reaching approximately $\sim 80\%$ for thin wires. Slightly lower values of approximately $\sim 60\text{--}70\%$ are obtained for the $C_4$ geometries.

{Our recent} 
proof-of-concept experiment 
for geometry I.6 in \cite{sakhno2025honeycomb} 
{demonstrated} a tunability of $64\%$ for a period-to-wire-radius ratio of $60$. The main source of discrepancy between {this} 
experimental result and the values shown in Fig.~\ref{fig:tunability}(a) (
{with} tunability exceeding $80\%$ for the same wire radius) is that the ideal minimal frequency configuration was not fully reached in the experimental setup. 


\section{Conclusion}

This work presents a general analytical framework for describing {tunability of} nested wire media with arbitrary lattice geometries. By extending the local field approach within the thin-wire approximation, the developed model enables 
rapid and efficient evaluation of dispersion characteristics for a wide class of complex structures, including geometries containing closely spaced wires through the proposed cluster 
approximation, without relying on time-consuming full-wave numerical simulations.

Our results demonstrate that the local field approach can be effectively used for the design of tunable wire media for plasma haloscope applications. This approach 
{predicts} high plasma frequency tunability {of} 
the proposed geometries via breathing deformation.

The analytical predictions were systematically validated by full-wave numerical simulations, demonstrating excellent agreement, particularly for thin-wire configurations. Our analysis of $C_6$ and $C_4$ symmetric geometries under breathing deformation reveals a remarkably high 
{tunability} of plasma frequency, 
ranging from $\sim 60\%$ to $80\%$. Notably, the $C_6$ configurations achieve a peak tunability of approximately $\sim 80\%$. These values significantly outperform previously reported mechanical tuning methods for wire media, 
{with typical} tuning range{s} of only $\sim 15\text{--}30\%$.



\newpage
\appendix

\section{Calculation of 
{i}nteraction constants} \label{app:formulae}

The interaction constant $D$ introduced in Eq.~(\ref{eq:ez_wm_r}) can be calculated using the Poisson summation formula. The result can be written as
\begin{align}
	&D (a,b,\theta; \boldsymbol{r};  k, \boldsymbol{q}) 
    =
    \label{eq:x_func_r}
    \\
    &-\frac{j \eta \kappa^2}{2 k}
	\sum \limits_{m=-\infty}^{+\infty}
	\Bigg[
	\frac{ e^{ +jq_x^{(m)} (\boldsymbol{r}, \hat{\boldsymbol{a}}) } }{k_x^{(m)} a} 
	e^{ 
		-j (
			(\boldsymbol{q}, \boldsymbol{b}) - q_x^{(m)} b_x
		) \, n^\prime 
	}
	\times
	\nonumber \\
	&
    \times
    \Bigg(
	\frac{
		\sin \left(
				k_x^{(m)} \left(
				b_y(1 - n^\prime)
				+ [\boldsymbol{r}, \hat{\boldsymbol{a}}]_z  
			\right) 
		\right)
	}{
		\cos \left( k_x^{(m)} b_y \right) - 
		\cos \left(
			q_x^{(m)} b_x - (\boldsymbol{q}, \boldsymbol{b})
		\right)
	}
    \nonumber \\
    & \qquad\quad
    +
    \frac{
		e^{+j 
			\left( 
			(\boldsymbol{q}, \boldsymbol{b}) - q_x^{(m)} b_x
			\right)
		}
		\sin \left(
		k_x^{(m)} \left(
		b_y n^\prime
		- [\boldsymbol{r}, \hat{\boldsymbol{a}}]_z 
		\right) 
		\right)
	}{
		\cos \left( k_x^{(m)} b_y \right) - 
		\cos \left(
			q_x^{(m)} b_x - (\boldsymbol{q}, \boldsymbol{b})
		\right)
	}
    \Bigg)
	\Bigg].
	\nonumber
\end{align}
where $q_{x}^{(m)} = 2\pi m/a + q_x$,  $k_{x}^{(m)}=-j\sqrt{(q_{x}^{(m)})^2-\kappa^2}$, $\boldsymbol{b}=(b_x, b_y)^\mathrm{T}$ ($b_x=b \cos \theta$, $b_y=b\sin \theta$){, and} 
$n^\prime$ is the minimal {integer} 
for which 
inequality $(nb_y - [\boldsymbol{r}, \hat{\boldsymbol{a}}]_z) \geq 0$ {is} satisfied. The component of the vector product  $[\boldsymbol{r}, \hat{\boldsymbol{a}}]_z = -r_y$.

The interaction constant $C$ defined right after Eq.~(\ref{eq:boundary_single_lat_2}) 
can be calculated using the Poisson summation formula
{, eliminating singularity similar to the calculation} 
for rectangular lattice in \cite{belov2002dispersion}
\begin{align}
	& C(a, b, \theta; k, \boldsymbol{q})=
    \label{eq:s_func} \\
    & -\frac{j \eta \kappa^2}{2k}
	\Bigg[
		\frac{1}{\pi} \left\{ \log \frac{\kappa a}{4 \pi} + \gamma \right\} - \frac{1}{2j}
		\nonumber \\
		& \qquad \qquad \qquad
		+ \frac{1}{k_x^{(0)}a}\,
		\frac{
			\sin \left( k_x^{(0)} b_y \right)
		}{
			\cos \left( k_x^{(0)} b_y \right) -
			\cos \left( q_y b_y \right)
		} 
		\nonumber \\
		& \quad
        + 
		\sum \limits_{m\neq 0}
		\Bigg(
        - 
		\frac{1}{2\pi |m|}
        + \frac{1}{k_x^{(m)}a} \times
        \nonumber \\
		& \qquad \qquad
		\times \frac{
			\sin \left( k_x^{(m)}b_y \right)
		}{
			\cos \left( k_x^{(m)} b_y \right) -
			\cos \left( q_x^{(m)} b_x - (\boldsymbol{q}, \boldsymbol{b}) \right)
		}
		\Bigg)
		\Bigg].
		\nonumber
\end{align}

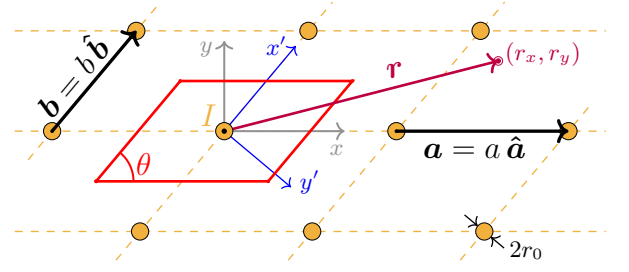
\begin{figure}
    \begin{minipage}{0.9\linewidth}
		\center{ 

\pgfset{
	foreach/parallel foreach/.style args={#1in#2via#3}{evaluate=#3 as #1 using {{#2}[#3-1]}},
}
\definecolor{gold}{rgb}{0.95, 0.69, 0.24}
\definecolor{grey}{rgb}{0.57, 0.57, 0.57}

\begin{tikzpicture}[scale=0.99, transform shape]
	\def \c{1.75}  
	\def \a{2.3}  
	\def \ang{50}  
	
	\def \shiftx {0.5 * \a + 0.5 * \c * cos(\ang)}
	\def \shifty {0.5 * \c * sin(\ang)}
	
	\coordinate(A1) at ({0 + \shiftx}, {0 + \shifty}); 
	\coordinate(A2) at ({\a + \shiftx}, {0 + \shifty}); 
	\coordinate(A3) at ({\a + \c * cos(\ang) + \shiftx}, {\c * sin(\ang) + \shifty}); 
	\coordinate(A4) at ({\c * cos(\ang) + \shiftx}, {\c * sin(\ang) + \shifty}); 
	\def \verts {A1, A2, A3, A4}
	
	\def \rw{0.05*\a}  
	\def \wirescol{gold}  
	
	\def \numa {2}  
	\def \numc {2}  
	\def \delta {0.5}  
	
	\foreach [evaluate={
		\linex = \c * cos(\ang) * \numc ;
		\liney = \c * sin(\ang) * \numc 
	}] \ia in {1, ..., \numa} {
	
		\coordinate(diag1) at ({ \ia * \a - \delta * cos(\ang) }, {- \delta * sin(\ang) });
		\coordinate(diag2) at ({ \ia * \a + \linex + \delta * cos(\ang) }, { \liney + \delta * sin(\ang) });
		
		\draw[dashed, line width=0.5pt, gold](diag1) -- (diag2);
	}
    \draw[dashed, line width=0.5pt, gold]
    ({ \c * cos(\ang) - \delta * cos(\ang) }, {\c * sin(\ang) - \delta * sin(\ang) }) -- ({2 * \c * cos(\ang) + \delta * cos(\ang) }, { 2 * \c * sin(\ang) + \delta * sin(\ang) });

    \draw[dashed, line width=0.5pt, gold]
    ({ 3 * \a - \delta * cos(\ang) }, { -\delta * sin(\ang) }) -- ({3 * \a + \c * cos(\ang) + \delta * cos(\ang) }, {\c * sin(\ang) + \delta * sin(\ang) });
	
	\foreach [evaluate={
		\rowy = \c * sin(\ang) * \ic
	}] \ic in {0, ..., \numc} {
		
		\coordinate(left) at ({ -\delta + \c * cos(\ang) }, { \rowy });
		\coordinate(right) at ({ 3 * \a + \c * cos(\ang) + \delta }, { \rowy });
		
		\draw[dashed, line width=0.5pt, gold](left) -- (right);
	}

	\def \refX {1 * \a + 1 * \c * cos(\ang)}
	\def \refY {1 * \c * sin(\ang)}
	
	\draw[->, line width=0.75pt, grey] ({\refX}, {\refY}) -- ({\refX + \a * 7 /10}, {\refY}) node[left, xshift=3pt, yshift=-7pt] {$x$};
	\draw[->, line width=0.75pt, grey] ({\refX}, {\refY}) -- ({\refX}, {\refY + \c * sin(\ang) * 9 / 10}) node[left, , xshift=-1pt, yshift=-3pt] {$y$};

    \node[
        gold, 
        xshift=-6pt, 
        yshift=6pt
    ] at ({\refX}, {\refY}) {\large $I$};

	\draw[->, line width=0.50pt, blue] ({\refX}, {\refY}) -- ({\refX + 8.5 / 10 * \c * cos(\ang)}, {\refY + 8.5 / 10 * \c * sin(\ang)}) node[left, xshift=0pt, yshift=0pt] {$x^\prime$};
	\draw[->, line width=0.50pt, blue] ({\refX}, {\refY}) -- ({\refX + 2 / 3 * \c * cos(90 - \ang)}, {\refY - 2 / 3 * \c * sin(90 - \ang)}) node[right, , xshift=0pt, yshift=2pt] {$y^\prime$};
    
	\draw[solid, line width=1.0pt, red](A1) -- (A2) -- (A3) -- (A4)--(A1);
	\foreach \vert in \verts {
		\filldraw [red] (\vert) circle (0.5pt); 
	}
	\filldraw [red] (A1) circle (0.5pt) node[right, xshift=12pt, yshift=6.5pt] {\large $\theta$};
	\draw pic[draw=orange, -, angle eccentricity=1.2, angle radius=0.5cm, line width=0.75pt, red]{angle=A2--A1--A4};

    \def \rcol{purple}  
    
	\coordinate (pntRef) at ({\refX}, {\refY});
	\coordinate (pntThis) at ({\refX + 1.25 * \a + 0.7 * \c * cos(\ang)}, {\refY + 0.7 * \c * sin(\ang});
	
	\filldraw [draw=\rcol, fill=white, line width=0.25pt] (pntThis) circle (\rw / 2);
	\filldraw [draw=\rcol, fill=\rcol, line width=0.25pt] (pntThis) circle (\rw / 4) node [above, xshift=17pt, yshift=-5pt, text=\rcol] {$(r_x, r_y)$};
		
	\draw[->, \rcol, line width=1.0pt] (pntRef) --(pntThis) node [midway, above, sloped, xshift=15pt, yshift=0pt] {\large $\boldsymbol{r}$};

	\def \ic {0}  
	\def \numa {3}  
	\foreach [evaluate={
		\thisx = \a * \ia + \c * cos(\ang) * \ic ;
		\thisy = \c * sin(\ang) * \ic 
	}] \ia in {1, ..., \numa} {
			\coordinate(this) at ({ \thisx }, { \thisy });
			\filldraw [black, fill=\wirescol, line width=0.25pt] (this){} circle (\rw);
	}
	
	\def \ic {1}  
	\def \numa {3}   
	\foreach [evaluate={
		\thisx = \a * \ia + \c * cos(\ang) * \ic ;
		\thisy = \c * sin(\ang) * \ic 
	}] \ia in {0, ..., \numa} {
		\coordinate(this) at ({ \thisx }, { \thisy });
		\filldraw [black, fill=\wirescol, line width=0.25pt] (this){} circle (\rw);
	}
	
	\def \ic {2}  
	\def \numa {2}    
	\foreach [evaluate={
		\thisx = \a * \ia + \c * cos(\ang) * \ic ;
		\thisy = \c * sin(\ang) * \ic 
	}] \ia in {0, ..., \numa} {
		\coordinate(this) at ({ \thisx }, { \thisy });
		\filldraw [black, fill=\wirescol, line width=0.25pt] (this){} circle (\rw);
	}
	
	
	\filldraw [black, fill=black, line width=0.25pt] ({\refX}, {\refY}) circle (\rw / 4);
	
	\def \circX {3 * \a + 0 * \c * cos(\ang)}
	\def \circY {0 * \c * sin(\ang)}
	\def \angArr {\ang + 90}

	\draw[->, black, line width=0.5pt]({\circX-3*\rw *cos(\angArr)}, {\circY-3*\rw * sin(\angArr)}) -- ({\circX-\rw*cos(\angArr)}, {\circY-\rw*sin(\angArr)}) node[midway,xshift=11pt, yshift=-3pt] {$2r_0$};
	\draw[->, black, line width=0.5pt]({\circX+3*\rw *cos(\angArr)}, {\circY+3*\rw * sin(\angArr)}) -- ({\circX+\rw*cos(\angArr)}, {\circY+\rw*sin(\angArr)});
	
	\coordinate (aLeft) at ({2 * \a + 1 * \c * cos(\ang)}, {1 * \c * sin(\ang});
	\coordinate (aRight) at ({\a + 2 * \a + 1 * \c * cos(\ang)}, {1 * \c * sin(\ang});
	\draw[->, line width=1.25pt, black] (aLeft) -- (aRight) node[midway, below, sloped, xshift=-3pt, yshift=2pt] {\large $\boldsymbol{a}=a\,  \boldsymbol{\hat a}$};
	\filldraw [black, fill=black, line width=0.25pt] (aLeft) circle (0.5pt);
    
	\coordinate (cLow) at ({0 * \a + 1 * \c * cos(\ang)}, {1 * \c * sin(\ang});
	\coordinate (cUp) at ({0 * \a + (1 + 1) * \c * cos(\ang)}, {(1 + 1) * \c * sin(\ang});
	\draw[->, line width=1.25pt, black] (cLow) -- (cUp) node[midway, above, sloped, xshift=-4pt, yshift=-1pt] {\large $\boldsymbol{b}=b\,  \boldsymbol{\hat b}$};
    \filldraw [black, fill=black, line width=0.25pt] (cLow) circle (0.5pt);
	
\end{tikzpicture}
		}
	\end{minipage}
	\caption{
		Geometry of a simple wire metamaterial formed by a single general lattice (with periods $a$ and $b$ {and} 
        an angle $\theta$ between lattice vectors) of parallel wires {with radius} 
        $r_0$. A unit cell is {indicated with red lines.} 
    } \label{fig:gen_cell_single} 
\end{figure}

{Calculating} 
the same interaction constants {with} 
the lattice summation (\ref{eq:green_x}) performed in different order gives alternative expressions for the interaction constants. This also can be interpreted as a {change of} coordinates 
from {the} $xy$ axes to {the} $x^\prime y^\prime$ axes plotted in Fig.~\ref{fig:gen_cell_single}. The change allows us to {obtain other} 
versions of expressions for $D$ and $C$ using {the} following substitutions in Eqs.~(\ref{eq:x_func_r}) and (\ref{eq:s_func}): (1) $\boldsymbol{a} \leftrightarrow \boldsymbol{b}$, (2) $(x,y,z) \rightarrow (x^\prime,y^\prime,z^\prime)$, $m \rightarrow n$.

{Substituting these expression into} 
Eq.~(\ref{eq:x_func_r}) gives:
\begin{align}
	&D (a,b,\theta; \boldsymbol{r};  k, \boldsymbol{q})
	= 
    \label{eq:x_func_another} \\
	& - \frac{\eta \kappa^2}{4 k} \, 2j
	\sum \limits_{n=-\infty}^{+\infty}
	\Bigg[
	\frac{ e^{ +j q_{x^\prime}^{(n)} (\boldsymbol{r}, \hat{\boldsymbol{b}}) } }{k_{x^\prime}^{(n)} b} 
	e^{ 
		-j \left(
		(\boldsymbol{q},\boldsymbol{a}) - q_{x^\prime}^{(n)}a_{x^\prime}
		\right) \, m^\prime
	} \times
	\nonumber \\
	&
    \times \Bigg(
	\frac{
		\sin \left(
		k_{x^\prime}^{(n)} \left(
		a_{y^\prime} (1 - m^\prime)
		+ [\boldsymbol{r}, \hat{\boldsymbol{b}} ]_{z^\prime}
		\right) 
		\right)
	}{
		\cos \left( k_{x^\prime}^{(n)} a_{y^\prime} \right) - 
		\cos \left(
		q_{x^\prime}^{(n)} a_{x^\prime} - (\boldsymbol{q},\boldsymbol{a})
		\right)
	} 
	\nonumber \\
	& \quad
	+
	\frac{
		e^{+j 
			\left( 
			(\boldsymbol{q},\boldsymbol{a}) - q_{x^\prime}^{(n)}a_{x^\prime}
			\right)
		}
		\sin \left(
		k_{x^\prime}^{(n)} \left(
		a_{y^\prime} m^\prime
		- [\boldsymbol{r}, \hat{\boldsymbol{b}} ]_{z^\prime}
		\right) 
		\right)
	}{
		\cos \left( k_{x^\prime}^{(n)} a_{y^\prime} \right) - 
		\cos \left(
		q_{x^\prime}^{(n)} a_{x^\prime} - (\boldsymbol{q},\boldsymbol{a})
		\right)
	}
	\Bigg)
	\Bigg],
	\nonumber
\end{align}
where  $q_{x^\prime}^{(n)} = 2\pi n / b + (\boldsymbol{q}, \hat{\boldsymbol{b}})=2\pi n / b + q_{x^\prime}${,} 
$k_{x^\prime}^{(n)}=-j\sqrt{(q_{x^\prime}^{(n)})^2-\kappa^2}${, and} 
$m^\prime$ is {the} 
minimal integer for which $(ma \, \sin \theta - [\boldsymbol{r}, \hat{\boldsymbol{b}} ]_{z^\prime}) \geq 0$. The vector product component $[\boldsymbol{r}, \hat{ \boldsymbol{b}} ]_{z^\prime}$ is equal to $-[\boldsymbol{r}, \hat{ \boldsymbol{b}} ]_z = -(r_x \sin\theta - r_y \cos\theta)=-r_{y^\prime}$ since {the} 
$z$ and $z^\prime$ {axes} are oppositely directed.

\begin{figure*}[t]
    \begin{minipage}{0.26\linewidth}
		\center{ 
	       \definecolor{gold}{rgb}{0.95, 0.69, 0.24}
\definecolor{grey}{rgb}{0.57, 0.57, 0.57}

\begin{tikzpicture}[scale=0.90, transform shape]
	\def \c{1.75}  
	\def \a{1.75}  
	\def \ang{60}  
    \def \rot {0}

    \def \rw {0.05 * \a}  

    \def \wiresDist {0.01 * \a}
    
	\def \RtoA {(0.5 - (\rw / \a) - (\wiresDist / \a))}

    \readlist*\wCols{gold, pink, brown, gold, pink, brown}
    
	\def \shiftx {0.5 * \a + 0.5 * \c * cos(\ang)}
	\def \shifty {0.5 * \c * sin(\ang)}
	
	\coordinate(A1) at ({0 + \shiftx}, {0 + \shifty}); 
	\coordinate(A2) at ({\a + \shiftx}, {0 + \shifty}); 
	\coordinate(A3) at ({\a + \c * cos(\ang) + \shiftx}, {\c * sin(\ang) + \shifty}); 
	\coordinate(A4) at ({\c * cos(\ang) + \shiftx}, {\c * sin(\ang) + \shifty}); 
	\def \verts {A1, A2, A3, A4}
	
    
	\def \wirescol{gold}  
    
	
	\def \numa {1}  
	\def \numc {2}  
	\def \delta {0.6}  
	\def \linesWidth {0.25pt}
    
	\foreach [evaluate={
		\linex = \c * cos(\ang) * \numc ;
		\liney = \c * sin(\ang) * \numc 
	}] \ia in {1, ..., \numa} {
	
		\coordinate(diag1) at ({ \ia * \a - \delta * cos(\ang) }, {- \delta * sin(\ang) });
		\coordinate(diag2) at ({ \ia * \a + \linex + \delta * cos(\ang) }, { \liney + \delta * sin(\ang) });
		
		\draw[dashed, line width=\linesWidth, gold](diag1) -- (diag2);
	}

    \foreach [evaluate={
		\linex = \c * cos(\ang) * \numc ;
		\liney = \c * sin(\ang) * \numc 
	}] \ia in {2} {
	
		\coordinate(diag1) at ({ \ia * \a + \delta * cos(\ang) }, {- \delta * sin(\ang) });
		\coordinate(diag2) at ({ \ia * \a - \linex - \delta * cos(\ang) }, { \liney + \delta * sin(\ang) });
		
		\draw[dashed, line width=\linesWidth, gold](diag1) -- (diag2);
	}
    
    \draw[dashed, line width=\linesWidth, gold] (
        {\c * cos(\ang) - \delta}, 
        {\c * sin(\ang) - \delta * tan(\ang) }
    ) -- (
        {2 * \c * cos(\ang) + \delta * cos(\ang) }, 
        {2 * \c * sin(\ang) + \delta * sin(\ang) }
    );

    \draw[dashed, line width=\linesWidth, gold] (
        { \c * cos(\ang) - \delta }, 
        {\c * sin(\ang) + \delta * tan(\ang) }
    ) -- (
        {\a + \delta * cos(\ang) },
        {- \delta * sin(\ang) }
    );

    \draw[dashed, line width=\linesWidth, gold] (
        {2 * \a - \delta * cos(\ang)}, 
        { -\delta * sin(\ang) }
    ) -- (
        {2 * \a + \c * cos(\ang) + \delta}, 
        {\c * sin(\ang) + \delta * tan(\ang)}
    );

    \draw[dashed, line width=\linesWidth, gold] (
        {2 * \a - \delta * cos(\ang)}, 
        {2 * \c * sin(\ang) + \delta * sin(\ang)}
    ) -- (
        {2 * \a + \c * cos(\ang) + \delta}, 
        {\c * sin(\ang) - \delta * tan(\ang)}
    );
    
	\foreach [evaluate={
		\rowy = \c * sin(\ang) * \ic
	}] \ic in {0, ..., \numc} {
		
		\coordinate(left) at ({ -\delta + \c * cos(\ang) }, { \rowy });
		\coordinate(right) at ({ 2 * \a + \c * cos(\ang) + \delta }, { \rowy });
		
		\draw[dashed, line width=\linesWidth, gold](left) -- (right);
	}

    
	\def \refX {1 * \a + 1 * \c * cos(\ang)}
	\def \refY {1 * \c * sin(\ang)}
	
    

    \def \halfD {\a / 2 / cos(30)};
    \def \angT {60};
    
    \def \hexCentX {\c * cos(\ang) + \a / 2}
    \def \hexCentY {\c * sin(\ang) + \a / 2 * sin(30) / cos(30)}
    
    \foreach \i in {0, ..., 5} {
	
		\coordinate(prev) at 
        (
        { \hexCentX + \halfD * cos((0.5 + \i) * \angT) }, 
        { \hexCentY + \halfD * sin((0.5 + \i) * \angT) }
        );
        
		\coordinate(this) at 
        (
        { \hexCentX + \halfD * cos((0.5 + \i + 1) * \angT) }, 
        { \hexCentY + \halfD * sin((0.5 + \i + 1) * \angT) }
        );

        \filldraw [black] (prev) circle (0.25pt); 
		\draw[solid, line width=0.75pt, black](prev) -- (this);
	}
    
    \def \hexCentX {\a + \c * cos(\ang)}
    \def \hexCentY {\c * sin(\ang)}
    
    \foreach \i in {0, ..., 5} {
	
		\coordinate(prev) at 
        (
        { \hexCentX + \halfD * cos((0.5 + \i) * \angT) }, 
        { \hexCentY + \halfD * sin((0.5 + \i) * \angT) }
        );
        
		\coordinate(this) at 
        (
        { \hexCentX + \halfD * cos((0.5 + \i + 1) * \angT) }, 
        { \hexCentY + \halfD * sin((0.5 + \i + 1) * \angT) }
        );

        \filldraw [pink] (prev) circle (0.1pt); 
		\draw[solid, line width=0.2pt, pink](prev) -- (this);
	}

    


    \def \ia {1}; \def \ic {1};
    \def \centerX {\a * \ia + \c * cos(\ang) * \ic};
    \def \centerY {\c * sin(\ang) * \ic};
    \foreach \i in {0, ..., 5} {
        \coordinate(this) at (
            { \centerX + \RtoA * \a * cos(\ang * \i + \rot)}, 
            { \centerY + \RtoA * \a * sin(\ang * \i + \rot)}
        );
        \filldraw [black, fill={\wCols[\i + 1]}, line width=0.25pt] (this) circle (\rw);
    }

    \def \ia {0}; \def \ic {1};
    \def \centerX {\a * \ia + \c * cos(\ang) * \ic};
    \def \centerY {\c * sin(\ang) * \ic};
    \foreach \i in {0, 1, 2, 4, 5} {
        \coordinate(this) at (
            { \centerX + \RtoA * \a * cos(\ang * \i + \rot)}, 
            { \centerY + \RtoA * \a * sin(\ang * \i + \rot)}
        );
        \filldraw [black, fill={\wCols[\i + 1]}, line width=0.25pt] (this) circle (\rw);
    }

    \def \ia {2}; \def \ic {1};
    \def \centerX {\a * \ia + \c * cos(\ang) * \ic};
    \def \centerY {\c * sin(\ang) * \ic};
    \foreach \i in {1, 2, 3, 4, 5} {
        \coordinate(this) at (
            { \centerX + \RtoA * \a * cos(\ang * \i + \rot)}, 
            { \centerY + \RtoA * \a * sin(\ang * \i + \rot)}
        );
        \filldraw [black, fill={\wCols[\i + 1]}, line width=0.25pt] (this) circle (\rw);
    }

    \foreach \ia in {0, 1} {
        \def \ic {2};
        \def \centerX {\a * \ia + \c * cos(\ang) * \ic};
        \def \centerY {\c * sin(\ang) * \ic};
        \foreach \i in {0, 3, 4, 5} {
            \coordinate(this) at (
                { \centerX + \RtoA * \a * cos(\ang * \i + \rot)}, 
                { \centerY + \RtoA * \a * sin(\ang * \i + \rot)}
            );
            \filldraw [black, fill={\wCols[\i + 1]}, line width=0.25pt] (this) circle (\rw);
        }
    }

    \foreach \ia in {1, 2} {
        \def \ic {0};
        \def \centerX {\a * \ia + \c * cos(\ang) * \ic};
        \def \centerY {\c * sin(\ang) * \ic};
        \foreach \i in {0, 1, 2, 3} {
            \coordinate(this) at (
                { \centerX + \RtoA * \a * cos(\ang * \i + \rot)}, 
                { \centerY + \RtoA * \a * sin(\ang * \i + \rot)}
            );
            \filldraw [black, fill={\wCols[\i + 1]}, line width=0.25pt] (this) circle (\rw);
        }
    }

    \def \ia {0}; \def \ic {0};
    \def \centerX {\a * \ia + \c * cos(\ang) * \ic};
    \def \centerY {\c * sin(\ang) * \ic};
    \foreach \i in {0, 1} {
        \coordinate(this) at (
            { \centerX + \RtoA * \a * cos(\ang * \i + \rot)}, 
            { \centerY + \RtoA * \a * sin(\ang * \i + \rot)}
        );
        \filldraw [black, fill={\wCols[\i + 1]}, line width=0.25pt] (this) circle (\rw);
    }

    \def \ia {-1}; \def \ic {2};
    \def \centerX {\a * \ia + \c * cos(\ang) * \ic};
    \def \centerY {\c * sin(\ang) * \ic};
    \foreach \i in {0, 5} {
        \coordinate(this) at (
            { \centerX + \RtoA * \a * cos(\ang * \i + \rot)}, 
            { \centerY + \RtoA * \a * sin(\ang * \i + \rot)}
        );
        \filldraw [black, fill={\wCols[\i + 1]}, line width=0.25pt] (this) circle (\rw);
    }

    \def \ia {2}; \def \ic {2};
    \def \centerX {\a * \ia + \c * cos(\ang) * \ic};
    \def \centerY {\c * sin(\ang) * \ic};
    \foreach \i in {3, 4} {
        \coordinate(this) at (
            { \centerX + \RtoA * \a * cos(\ang * \i + \rot)}, 
            { \centerY + \RtoA * \a * sin(\ang * \i + \rot)}
        );
        \filldraw [black, fill={\wCols[\i + 1]}, line width=0.25pt] (this) circle (\rw);
    }

    \def \ia {3}; \def \ic {0};
    \def \centerX {\a * \ia + \c * cos(\ang) * \ic};
    \def \centerY {\c * sin(\ang) * \ic};
    \foreach \i in {2, 3} {
        \coordinate(this) at (
            { \centerX + \RtoA * \a * cos(\ang * \i + \rot)}, 
            { \centerY + \RtoA * \a * sin(\ang * \i + \rot)}
        );
        \filldraw [black, fill={\wCols[\i + 1]}, line width=0.25pt] (this) circle (\rw);
    }
	



\end{tikzpicture}
		}
	\end{minipage}
    \put(-133, 60){(a)}
    \put(-113, 60){I.6}
    \hfill
	\begin{minipage}{0.26\linewidth}
		\center{
            \input{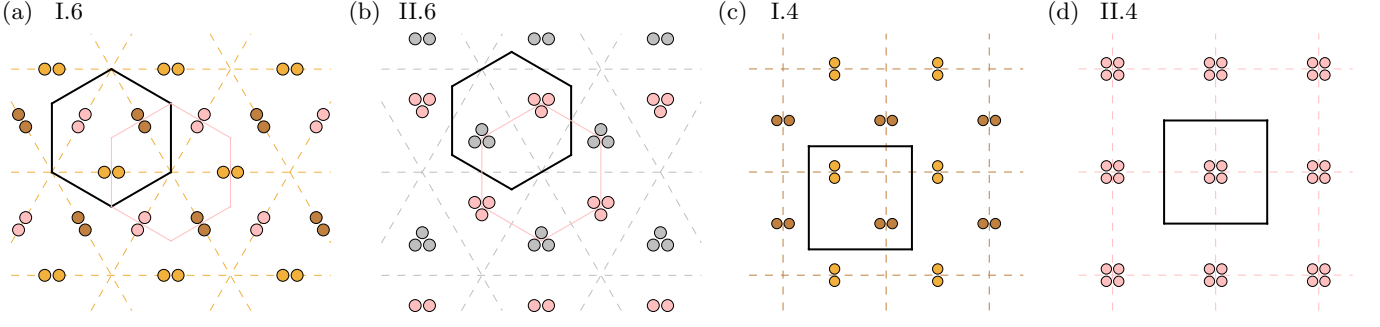}
		}
	\end{minipage}
    \put(-142, 60){(b)}
    \put(-123, 60){II.6}
    \hfill
	\begin{minipage}{0.21\linewidth}
		\center{
            \definecolor{gold}{rgb}{0.95, 0.69, 0.24}
\definecolor{grey}{rgb}{0.57, 0.57, 0.57}

\begin{tikzpicture}[scale=0.90, transform shape]
	\def \c{(1.75 * sin(60))}  
	\def \a{(1.75 * sin(60))}  
	\def \ang{90}  
    \def \rot {0}

    \def \rw {0.05 * \a}  

    \def \wiresDist {0.01 * \a}
    
	\def \RtoA {(0.5 - (\rw / \a) - (\wiresDist / \a))}

    \readlist*\wCols{brown, gold, brown, gold}
    
	\def \shiftx {0.25 * \a}
	\def \shifty {0.25 * \c}
	
	\coordinate(A1) at ({0 + \shiftx}, {0 + \shifty}); 
	\coordinate(A2) at ({\a + \shiftx}, {0 + \shifty}); 
	\coordinate(A3) at ({\a + \c * cos(\ang) + \shiftx}, {\c * sin(\ang) + \shifty}); 
	\coordinate(A4) at ({\c * cos(\ang) + \shiftx}, {\c * sin(\ang) + \shifty}); 
	\def \verts {A1, A2, A3, A4}
	
    
	\def \wirescol{brown}  
    
	
	\def \numa {2}  
	\def \numc {2}  
	\def \delta {0.5}  
    \def \deltaY {0.6 * sin(60)}
	\def \linesWidth {0.25pt}
    
	\foreach [evaluate={
		\linex = \c * cos(\ang) * \numc ;
		\liney = \c * sin(\ang) * \numc 
	}] \ia in {0, ..., \numa} {
	
		\coordinate(diag1) at ({ \ia * \a - \deltaY * cos(\ang) }, {- \deltaY * sin(\ang) });
		\coordinate(diag2) at ({ \ia * \a + \linex + \deltaY * cos(\ang) }, { \liney + \deltaY * sin(\ang) });
		
		\draw[dashed, line width=\linesWidth, \wirescol](diag1) -- (diag2);
	}
    
	\foreach [evaluate={
		\rowy = \c * sin(\ang) * \ic
	}] \ic in {0, ..., \numc} {
		
		\coordinate(left) at ({ -\delta + \c * cos(\ang) }, { \rowy });
		\coordinate(right) at ({ 2 * \a + \c * cos(\ang) + \delta }, { \rowy });
		
		\draw[dashed, line width=\linesWidth, \wirescol](left) -- (right);
	}

    \coordinate(left) at ({ -\delta + \c * cos(\ang) }, { \c * sin(\ang) * 2.61 });
    \coordinate(right) at ({ 2 * \a + \c * cos(\ang) + \delta }, { \c * sin(\ang) * 2.61 });
    
    \draw[dashed, line width=\linesWidth, white](left) -- (right);

    
	\def \refX {1 * \a + 1 * \c * cos(\ang)}
	\def \refY {1 * \c * sin(\ang)}
	
    
	\draw[solid, line width=0.75pt, black](A1) -- (A2) -- (A3) -- (A4)--(A1);
	\foreach \vert in \verts {
		\filldraw [black] (\vert) circle (0.25pt); 
	}

    


    \foreach \ia in {0, 1} {
        \foreach \ic in {0, 1} {
            \def \centerX {\a / 2 + \a * \ia};
            \def \centerY {\c / 2 + \c * \ic};
            \foreach \i in {0, ..., 3} {
                \coordinate(this) at (
                    { \centerX + \RtoA * \a * cos(\ang * \i + \rot)}, 
                    { \centerY + \RtoA * \a * sin(\ang * \i + \rot)}
                );
                \filldraw [black, fill={\wCols[\i + 1]}, line width=0.25pt] (this) circle (\rw);
            }
        }
    }

    \foreach \ia in {0, 1} {
        \def \ic {-1};
        \def \centerX {\a / 2 + \a * \ia};
        \def \centerY {\c / 2 + \c * \ic};
        \foreach \i in {1} {
            \coordinate(this) at (
                { \centerX + \RtoA * \a * cos(\ang * \i + \rot)}, 
                { \centerY + \RtoA * \a * sin(\ang * \i + \rot)}
            );
            \filldraw [black, fill={\wCols[\i + 1]}, line width=0.25pt] (this) circle (\rw);
        }
    }

    \foreach \ia in {0, 1} {
        \def \ic {2};
        \def \centerX {\a / 2 + \a * \ia};
        \def \centerY {\c / 2 + \c * \ic};
        \foreach \i in {3} {
            \coordinate(this) at (
                { \centerX + \RtoA * \a * cos(\ang * \i + \rot)}, 
                { \centerY + \RtoA * \a * sin(\ang * \i + \rot)}
            );
            \filldraw [black, fill={\wCols[\i + 1]}, line width=0.25pt] (this) circle (\rw);
        }
    }

    \foreach \ic in {0, 1} {
        \def \ia {-1};
        \def \centerX {\a / 2 + \a * \ia};
        \def \centerY {\c / 2 + \c * \ic};
        \foreach \i in {0} {
            \coordinate(this) at (
                { \centerX + \RtoA * \a * cos(\ang * \i + \rot)}, 
                { \centerY + \RtoA * \a * sin(\ang * \i + \rot)}
            );
            \filldraw [black, fill={\wCols[\i + 1]}, line width=0.25pt] (this) circle (\rw);
        }
    }

    \foreach \ic in {0, 1} {
        \def \ia {2};
        \def \centerX {\a / 2 + \a * \ia};
        \def \centerY {\c / 2 + \c * \ic};
        \foreach \i in {2} {
            \coordinate(this) at (
                { \centerX + \RtoA * \a * cos(\ang * \i + \rot)}, 
                { \centerY + \RtoA * \a * sin(\ang * \i + \rot)}
            );
            \filldraw [black, fill={\wCols[\i + 1]}, line width=0.25pt] (this) circle (\rw);
        }
    }
    

	

	
\end{tikzpicture}
		}
	\end{minipage}
    \put(-117, 60){(c)}
    \put(-97, 60){I.4}
    \hfill
    \hspace{7pt}
	\begin{minipage}{0.21\linewidth}
		\center{
            \definecolor{gold}{rgb}{0.95, 0.69, 0.24}
\definecolor{grey}{rgb}{0.57, 0.57, 0.57}

\begin{tikzpicture}[scale=0.90, transform shape]
	\def \c{(1.75 * sin(60))}  
	\def \a{(1.75 * sin(60))}  
	\def \ang{90}  
    \def \rot {45}

    \def \rw {0.05 * \a}  

    \def \wiresDist {0.01 * \a}
    
	\def \RtoA {( (0.5 - (\rw / \a) - (\wiresDist / \a)) * sqrt(2) )}

    \readlist*\wCols{pink, pink, pink, pink}
    
	\def \shiftx {0.5 * \a}
	\def \shifty {0.5 * \a}
	
	\coordinate(A1) at ({0 + \shiftx}, {0 + \shifty}); 
	\coordinate(A2) at ({\a + \shiftx}, {0 + \shifty}); 
	\coordinate(A3) at ({\a + \c * cos(\ang) + \shiftx}, {\c * sin(\ang) + \shifty}); 
	\coordinate(A4) at ({\c * cos(\ang) + \shiftx}, {\c * sin(\ang) + \shifty}); 
	\def \verts {A1, A2, A3, A4}
	
    
	\def \wirescol{pink}  
    
	
	\def \numa {2}  
	\def \numc {2}  
	\def \delta {0.5}  
    \def \deltaY {0.6 * sin(60)}
	\def \linesWidth {0.25pt}
    
	\foreach [evaluate={
		\linex = \c * cos(\ang) * \numc ;
		\liney = \c * sin(\ang) * \numc 
	}] \ia in {0, ..., \numa} {
	
		\coordinate(diag1) at ({ \ia * \a - \deltaY * cos(\ang) }, {- \deltaY * sin(\ang) });
		\coordinate(diag2) at ({ \ia * \a + \linex + \deltaY * cos(\ang) }, { \liney + \deltaY * sin(\ang) });
		
		\draw[dashed, line width=\linesWidth, \wirescol](diag1) -- (diag2);
	}
    
	\foreach [evaluate={
		\rowy = \c * sin(\ang) * \ic
	}] \ic in {0, ..., \numc} {
		
		\coordinate(left) at ({ -\delta + \c * cos(\ang) }, { \rowy });
		\coordinate(right) at ({ 2 * \a + \c * cos(\ang) + \delta }, { \rowy });
		
		\draw[dashed, line width=\linesWidth, \wirescol](left) -- (right);
	}

    \coordinate(left) at ({ -\delta + \c * cos(\ang) }, { \c * sin(\ang) * 2.61 });
    \coordinate(right) at ({ 2 * \a + \c * cos(\ang) + \delta }, { \c * sin(\ang) * 2.61 });
    
    \draw[dashed, line width=\linesWidth, white](left) -- (right);

    
	\def \refX {1 * \a + 1 * \c * cos(\ang)}
	\def \refY {1 * \c * sin(\ang)}
	
    
	\draw[solid, line width=0.75pt, black](A1) -- (A2) -- (A3) -- (A4)--(A1);
	\foreach \vert in \verts {
		\filldraw [black] (\vert) circle (0.25pt); 
	}

    


    \foreach \ia in {0, 1} {
        \foreach \ic in {0, 1} {
            \def \centerX {\a / 2 + \a * \ia};
            \def \centerY {\c / 2 + \c * \ic};
            \foreach \i in {0, ..., 3} {
                \coordinate(this) at (
                    { \centerX + \RtoA * \a * cos(\ang * \i + \rot)}, 
                    { \centerY + \RtoA * \a * sin(\ang * \i + \rot)}
                );
                \filldraw [black, fill={\wCols[\i + 1]}, line width=0.25pt] (this) circle (\rw);
            }
        }
    }

    \foreach \ia in {0, 1} {
        \def \ic {-1};
        \def \centerX {\a / 2 + \a * \ia};
        \def \centerY {\c / 2 + \c * \ic};
        \foreach \i in {0, 1} {
            \coordinate(this) at (
                { \centerX + \RtoA * \a * cos(\ang * \i + \rot)}, 
                { \centerY + \RtoA * \a * sin(\ang * \i + \rot)}
            );
            \filldraw [black, fill={\wCols[\i + 1]}, line width=0.25pt] (this) circle (\rw);
        }
    }

    \foreach \ia in {0, 1} {
        \def \ic {2};
        \def \centerX {\a / 2 + \a * \ia};
        \def \centerY {\c / 2 + \c * \ic};
        \foreach \i in {2, 3} {
            \coordinate(this) at (
                { \centerX + \RtoA * \a * cos(\ang * \i + \rot)}, 
                { \centerY + \RtoA * \a * sin(\ang * \i + \rot)}
            );
            \filldraw [black, fill={\wCols[\i + 1]}, line width=0.25pt] (this) circle (\rw);
        }
    }

    \foreach \ic in {0, 1} {
        \def \ia {-1};
        \def \centerX {\a / 2 + \a * \ia};
        \def \centerY {\c / 2 + \c * \ic};
        \foreach \i in {0, 3} {
            \coordinate(this) at (
                { \centerX + \RtoA * \a * cos(\ang * \i + \rot)}, 
                { \centerY + \RtoA * \a * sin(\ang * \i + \rot)}
            );
            \filldraw [black, fill={\wCols[\i + 1]}, line width=0.25pt] (this) circle (\rw);
        }
    }

    \def \ia {-1}; \def \ic {-1};
    \def \centerX {\a / 2 + \a * \ia};
    \def \centerY {\c / 2 + \c * \ic};
    \foreach \i in {0} {
        \coordinate(this) at (
            { \centerX + \RtoA * \a * cos(\ang * \i + \rot)}, 
            { \centerY + \RtoA * \a * sin(\ang * \i + \rot)}
        );
        \filldraw [black, fill={\wCols[\i + 1]}, line width=0.25pt] (this) circle (\rw);
    }

    \def \ia {-1}; \def \ic {2};
    \def \centerX {\a / 2 + \a * \ia};
    \def \centerY {\c / 2 + \c * \ic};
    \foreach \i in {3} {
        \coordinate(this) at (
            { \centerX + \RtoA * \a * cos(\ang * \i + \rot)}, 
            { \centerY + \RtoA * \a * sin(\ang * \i + \rot)}
        );
        \filldraw [black, fill={\wCols[\i + 1]}, line width=0.25pt] (this) circle (\rw);
    }

    \foreach \ic in {0, 1} {
        \def \ia {2};
        \def \centerX {\a / 2 + \a * \ia};
        \def \centerY {\c / 2 + \c * \ic};
        \foreach \i in {1, 2} {
            \coordinate(this) at (
                { \centerX + \RtoA * \a * cos(\ang * \i + \rot)}, 
                { \centerY + \RtoA * \a * sin(\ang * \i + \rot)}
            );
            \filldraw [black, fill={\wCols[\i + 1]}, line width=0.25pt] (this) circle (\rw);
        }
    }

    \def \ia {2}; \def \ic {-1};
    \def \centerX {\a / 2 + \a * \ia};
    \def \centerY {\c / 2 + \c * \ic};
    \foreach \i in {1} {
        \coordinate(this) at (
            { \centerX + \RtoA * \a * cos(\ang * \i + \rot)}, 
            { \centerY + \RtoA * \a * sin(\ang * \i + \rot)}
        );
        \filldraw [black, fill={\wCols[\i + 1]}, line width=0.25pt] (this) circle (\rw);
    }

    \def \ia {2}; \def \ic {2};
    \def \centerX {\a / 2 + \a * \ia};
    \def \centerY {\c / 2 + \c * \ic};
    \foreach \i in {2} {
        \coordinate(this) at (
            { \centerX + \RtoA * \a * cos(\ang * \i + \rot)}, 
            { \centerY + \RtoA * \a * sin(\ang * \i + \rot)}
        );
        \filldraw [black, fill={\wCols[\i + 1]}, line width=0.25pt] (this) circle (\rw);
    }
    

	


\end{tikzpicture}
		}
	\end{minipage}
    \put(-117, 60){(d)}
    \put(-97, 60){II.4}
	\caption{
		Limit{ing} cases {of touching wires, corresponding} 
        to the plasma frequency 
        $k_p^\text{lim}$ (see Fig.~\ref{fig:freq_tuning}). Wires are clustered, all wires from one cluster are drawn {in the} 
        same color.
	} \label{fig:geoms_lim} 
\end{figure*}

\begin{table*}[t]
  \centering
  \begin{tabular}{| c|| c||c|c|| c||c|c|| c||c|c ||}
    \hline \rule{0pt}{9pt}
    \multirow{3}{*}{$a/r_0$} 
    & \multicolumn{3}{c||}{ \textbf{I.6} } 
    & \multicolumn{3}{c||}{ \textbf{II.6} } 
    & \multicolumn{3}{c||}{ \textbf{I.4} }\\
    \cline{2-10}
    \rule{0pt}{9pt}
    & \multicolumn{1}{c||}{ numerical } 
    & \multicolumn{2}{c||}{ analytical } 
    & \multicolumn{1}{c||}{ numerical } 
    & \multicolumn{2}{c||}{ analytical } 
    & \multicolumn{1}{c||}{ numerical } 
    & \multicolumn{2}{c||}{ analytical } 
    \\
    \cline{2-10}
    \rule{0pt}{9pt}
    & $k_p^\text{lim}$ & Eq. (\ref{eq:system_clusters}) &
    $\Delta k_p^\text{lim}(\%)$
    & $k_p^\text{lim}$ & Eq. (\ref{eq:system_clusters}) &
    $\Delta k_p^\text{lim}(\%)$
    & $k_p^\text{lim}$ & Eq. (\ref{eq:system_clusters}) &
    $\Delta k_p^\text{lim}(\%)$
    \\ 
    \hline
    \hline
    20 & 
    0.7673 & 0.7165 & $6.62$ &  
    0.6301 & 0.5816 & $7.71$ &  
    0.5536 & 0.5279 & $4.64$    
    \\
    25 & 
    0.7003 & 0.6654 & $4.98$ &  
    0.5746 & 0.5377 & $6.42$ &  
    0.5069 & 0.4873 & $3.87$   
    \\
    50 & 
    0.5620 & 0.5460 & $2.85$ &  
    0.4599 & 0.4402 & $4.28$ &  
    0.4106 & 0.3999 & $2.61$   
    \\
    100 & 
    0.4793 & 0.4694 & $2.07$&  
    0.3917 & 0.3791 & $3.22$&  
    0.3527 & 0.3458 & $1.96$   
    \\
    200 & 
    0.4238 & 0.4169 & $1.63$&  
    0.3461 & 0.3372 & $2.57$&  
    0.3135 & 0.3086 & $1.56$   
    \\
    1000 & 
    0.3443 & 0.3405 & $1.10$&  
    0.2810 & 0.2761 & $1.74$ &  
    0.2567 & 0.2540 & $1.05$   
    \\
    \hline
  \end{tabular}
  \caption{
  Table of $k_p^\text{lim}$ values ($2 \pi/a$ units) for geometries from Fig.~\ref{fig:geoms} obtained numerically and analytically via Eq.~(\ref{eq:system_clusters}). For the geometry II.4{,} $k_p^\text{lim}$ is equal to $k_p^\text{min}$ (see Fig.~\ref{fig:deforms}(d)) from {T}able \ref{tab:estimations_min}. \\
  }
  \label{tab:estimations_lim}
\end{table*}

Equation (\ref{eq:s_func}) transforms into
\begin{align}
	& C(a, b, \theta; k, \boldsymbol{q}) = 
    \label{eq:s_func_another} \\
    &-\frac{\eta \kappa^2}{4k} \, 2j
	\Bigg[
	\frac{1}{\pi} \left\{ \log \frac{\kappa b}{4 \pi} + \gamma \right\}
	- \frac{1}{2j}
	\nonumber \\
	& \qquad \qquad \qquad
	+ \frac{1}{k_{x^\prime}^{(0)} b}\,
	\frac{
		\sin \left( k_{x^\prime}^{(0)} a_{y^\prime} \right)
	}{
		\cos \left( k_{x^\prime}^{(0)} a_{y^\prime} \right) -
		\cos \left( q_{y^\prime}^{(0)} a_{y^\prime} \right)
	} 
	\nonumber \\
	& \quad
    +
	\sum \limits_{n\neq 0}
	\Bigg(
    - \frac{1}{2\pi |n|}
    +
	\frac{1}{k_{x^\prime}^{(n)} b} \times
    \nonumber \\
    & \qquad \qquad
    \times
	\frac{
		\sin \left( k_{x^\prime}^{(m)}a_{y^\prime} \right)
	}{
		\cos \left( k_{x^\prime}^{(n)} a_{y^\prime} \right) -
		\cos \left( q_{x^\prime}^{(n)} a_{x^\prime} - (\boldsymbol{q}, \boldsymbol{a}) \right)
	}
	\Bigg)
	\Bigg]
	.\nonumber
\end{align}

{In the $x^\prime y^\prime$ axes,} the lattice vector $\boldsymbol{a}$ has the 
coordinates 
$(a_{x^\prime}, a_{y^\prime})^\mathrm{T}=(a\cos \theta, a\sin\theta)^\mathrm{T}$. The dot product $(\boldsymbol{a}, \boldsymbol{q})$ is invariant and can be calculated either in {the} $xy$ or $x^\prime y^\prime$ axes.

The 
calculation method (\ref{eq:x_func_r} or \ref{eq:x_func_another}, \ref{eq:s_func} or \ref{eq:s_func_another}) for each interaction constant can be {selected} 
based on their convergence rates. Since both analytical forms yield identical numerical results, any form can be used at any position within the boundary condition {defined by the} 
matrix (\ref{eq:system_n}).

For example, at the $\Gamma$-point, the expression (\ref{eq:x_func_r}) exhibits poor convergence along the lines parallel to the $\hat{\boldsymbol{a}}$ direction in the $xy$-plane, defined by $\boldsymbol{r}^{(1)} = x\hat{\boldsymbol{a}} + p\boldsymbol{b}$ (where $p$ is an integer; at these points, $n^\prime b_y - [\boldsymbol{r}^{(1)},\hat{\boldsymbol{a}}]_z = 0$). Conversely, the series in 
(\ref{eq:x_func_another}) 
{converges poorly} along the lines parallel to the $\hat{\boldsymbol{b}}$ direction, defined by $\boldsymbol{r}^{(2)} = l \boldsymbol{a} + y \hat{\boldsymbol{b}}$ (where $l$ is an integer; at these points, $m^\prime a_y - [\boldsymbol{r}^{(2)},\hat{\boldsymbol{b}}]_{z^\prime} = 0$).

Consequently, one can use 
(\ref{eq:x_func_another}) instead of (\ref{eq:x_func_r}) along the lines $\boldsymbol{r}^{(1)}$, and vice versa, 
(\ref{eq:x_func_r}) instead of (\ref{eq:x_func_another}) along the lines $\boldsymbol{r}^{(2)}$. In the immediate vicinity of the lattice points $\boldsymbol{R}^{(l,p)}$ (the intersection{s} 
of the $\boldsymbol{r}^{(1)}$ and $\boldsymbol{r}^{(2)}$ lines), both expressions converge slowly. To address this, a dedicated small-argument approximation {(\ref{eq:x_close})} for the interaction constant $D$ is proposed in Appendix \ref{app:clusters}. 

At the $\Gamma$-point, the two analytical forms of the interaction constant $C$ are mathematically identical when the lattice periods are equal ($a=b$). For asymmetric lattices {with} 
$a<b$, the expression (\ref{eq:s_func}) {converges faster.} 
Conversely, when $a>b$, the formula (\ref{eq:s_func_another}) converges more rapidly and should be used instead {of \ref{eq:s_func}}.

\section{Cluster 
approximation} \label{app:clusters}

\begin{figure}[h!]
    \begin{minipage}{0.9\linewidth}
		\center{ 
			\input{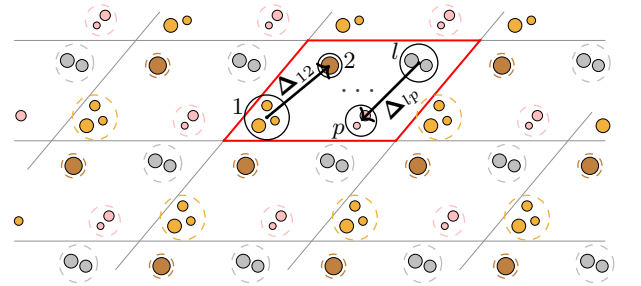}
		}
    \end{minipage}
	\caption{
		Example of {a} \textit{clustered} wire medi{um, in which close} 
        wires can be grouped in clusters. 
    } \label{fig:gen_cell_clusters} 
\end{figure}

In this part{,} we discuss {how to simplify} 
the matrix $\mathbb{M}$ 
if some wires are located close to each other. 
Fig{ure} \ref{fig:gen_cell_clusters} {illustrates such a wire medium, with some wire groups} 
highlighted. 
Within these groups{,} we assume that {any two wires are closely located, i.e.~the} 
vector $\Delta^{(p)}_{mn}$ between {the} $m^\text{th}$ and $n^\text{th}$ wires belonging to {the} $p^\text{th}$ cluster 
{is} small, 
$\Delta^{(p)}_{mn} \ll a,b$.

\begin{table*}[t]
  \centering
  \begin{tabular}{| c || c||c|c|| c|c||c|c|c||}
    \hline \rule{0pt}{9pt}
    \multirow{3}{*}{$a/r_0$} & 
    \multicolumn{3}{c||}{ \textbf{I.6} ($R/a=1/3$) } & 
    \multicolumn{5}{c||}{ \textbf{II.6} }
    \\
    \cline{2-9}
    \rule{0pt}{9pt}
    & \multicolumn{1}{c||}{ numerical } &
    \multicolumn{2}{c||}{ analytical } & 
    \multicolumn{2}{c||}{ numerical } &
    \multicolumn{3}{c||}{ analytical } 
    \\
    \cline{2-9}
    \rule{0pt}{9pt}
    & 
    $k_p^\text{max}$ & 
    $k_p^\text{max}$ from Eq. (\ref{eq:det_N}) &
    $\Delta k_p^\text{max}(\%)$
    & $R/a$
    & $k_p^\text{max}$ 
    & $R/a$ & $k_p^\text{max}$ from Eq. (\ref{eq:det_N}) &
    $\Delta k_p^\text{max}(\%)$
    \\ 
    \hline
    \hline
    20 & 
    1.0518 & 1.0409 & 
    1.036 &
    0.3314 & 1.0824 &  
    0.3316 & 1.0696 & 
    1.183 
    \\
    25 &
    0.9645 & 0.9597 & 
    0.498 &  
    0.3331 & 0.9867 &  
    0.3332 & 0.9812 & 
    0.557  
    \\
    50 &
    0.7798 & 0.7794 & 
    0.051 & 
    0.3359 & 0.7900 &  
    0.3360 & 0.7896 & 
    0.051  
    \\
    100 &
    0.6677 & 0.6677 & 
    0.005&
    0.3373 & 0.6736 &  
    0.3372 & 0.6736 & 
    0.004  
    \\
    200 &
    0.5919 & 0.5919 & 
    0.002&
    0.3378 & 0.5959 &  
    0.3379 & 0.5959 & 
    0.001
    \\
    1000 &
    0.4826 & 0.4826 & 
    $<$0.001&  
    0.3387 & 0.4847 &  
    0.3386 & 0.4847 & 
    $<$0.001 
    \\
    \hline
  \end{tabular}
  \caption{
  Table of $k_p^\text{max}$ values ($2 \pi/a$ units) for hexagonal geometries I.6 and II.6 obtained numerically and analytically via Eq.~(\ref{eq:det_N}).
  }
  \label{tab:estimations_max_hex}
\end{table*}
\begin{table*}[t]
  \centering
  \begin{tabular}{| c || c|c||c|c|c || c||c|c ||}
    \hline \rule{0pt}{9pt}
    \multirow{3}{*}{$a/r_0$} & 
    \multicolumn{5}{c||}{ \textbf{I.4} } & 
    \multicolumn{3}{c||}{ \textbf{II.4} ($R/a=1/2\sqrt{2}$) }
    \\
    \cline{2-9}
    \rule{0pt}{9pt}
    & 
    \multicolumn{2}{c||}{ numerical } &
    \multicolumn{3}{c||}{ analytical } 
    & \multicolumn{1}{c||}{ numerical } &
    \multicolumn{2}{c||}{ analytical }
    \\
    \cline{2-9}
    \rule{0pt}{9pt}
    & $R/a$
    & $k_p^\text{max}$ 
    & $R/a$ & $k_p^\text{max}$ from Eq. (\ref{eq:det_N}) &
    $\Delta k_p^\text{max}(\%)$
    & 
    $k_p^\text{max}$ & 
    $k_p^\text{max}$ from Eq. (\ref{eq:det_N}) &
    $\Delta k_p^\text{max}(\%)$
    \\ 
    \hline
    \hline
    20 & 
    0.3334 & 0.6657 &   
    0.3316 & 0.6635 &  
    0.330 &
    0.7506 & 0.7467 &
    0.520
    \\
    25 &
    0.3311 & 0.6268 &   
    0.3311 & 0.6258 &  
    0.160 &
    0.6908 & 0.6891 &
    0.246 
    \\
    50 &
    0.3273 & 0.5336 &   
    0.3272 & 0.5335 &  
    0.019 &
    0.5658 & 0.5656 &
    0.035
    \\
    100 &
    0.3254 & 0.4696 &   
    0.3253 & 0.4696 &  
    0.002 &
    0.4891 & 0.4891 &
    0.003 
    \\
    200 &
    0.3243 & 0.4232 &   
    0.3242 & 0.4232 &  
    $<$0.001 &
    0.4365 & 0.4365 &
    0.001 
    \\
    1000 &
    0.3228 & 0.3522 &   
    0.3229 & 0.3522 &  
    $<$0.001 &
    0.3591 & 0.3591 &
    $<$0.001
    \\
    \hline
  \end{tabular}
  \caption{
  Table of $k_p^\text{max}$ values ($2 \pi/a$ units) for square geometries I.4 and II.4 obtained numerically and analytically via Eq.~(\ref{eq:det_N}).
  }
  \label{tab:estimations_max_sqr}
\end{table*}

Under this assumption, we 
write an approximation for the interaction constant $D$ using the interaction constant $C$:
\begin{align}
    &D (a,b,\theta, \boldsymbol{\Delta}^{(p)}_{mn};k, \boldsymbol{q})
    \approx
    \nonumber \\
    &D_\text{close}
    (a,b,\theta, \boldsymbol{\Delta}^{(p)}_{mn};k, \boldsymbol{q}) 
    =
    C(a,b,\theta; k, \boldsymbol{q}) 
    \label{eq:x_close}
    \\
    & \qquad\qquad
    -\frac{\eta \kappa^2}{4k}
    \left\{
        1 - \frac{2j}{\pi} \left( \log \frac{\kappa \Delta^{(p)}_{mn}}{2} + \gamma \right)
    \right\} e^{-j (\boldsymbol{q}, \boldsymbol{\Delta}^{(p)}_{mn})}
    \nonumber
\end{align}
Hence, for each $p^\text{th}$ cluster, which consists of $N_p$ wires of radii $\{r_n\}_{n=1}^{N_p}${,} we can write \textit{a cluster interaction matrix}:
\begin{align}
    \mathbb{M}_{nm}^{\text{($p$)}} = 
    & \delta_{nm}
    \left[   
    \alpha^{-1}(r_n;k,q_z)
    - C(a,b,\theta;k,\boldsymbol{q})
    \right]
    \nonumber
    \\
    - 
    (1 - & \delta_{nm}) 
    D_\text{close}
    (a,b,\theta; \boldsymbol{\Delta}_{mn}^{(p)}; k,\boldsymbol{q}),
    \label{eq:cluster_matrix}
\end{align}
where $\delta_{nm}$ is the Kronecker delta.
Therefore, to calculate the cluster matrix{, we need to calculate only} the 
interaction constant $C$ 
instead of $N_p\times (N_p - 1)$ $D$ constants together with the same $C$ constant for the matrix of {the} complete system (\ref{eq:system_n}). This approximation {significantly accelerates the} 
calculations. 

{At the $\Gamma$-point ($\boldsymbol{q} = \boldsymbol{0}$),} the matrix (\ref{eq:cluster_matrix}) can be simplified 
{by substituting} $\alpha^{-1}$ from Eq.~(\ref{eq:eff_suscept}):
\begin{align}
    &\mathbb{M}_{nm}^{\text{($p$)}}(\boldsymbol{q}=\boldsymbol{0}) = 
    \label{eq:cluster_matrix_gamma}
    \\
    &
    \left[
    C(a,b,\theta;k, \boldsymbol{0})
    -
    \frac{\eta k}{4} 
    \left\{ 
    1 - \frac{2j}{\pi}
    \left(
        \log \frac{\kappa a}{2} + \gamma 
    \right)
    \right\}
    \right]
    \nonumber
    \\
    & + 
    \frac{\eta k}{4} 
    \frac{2j}{\pi}
    \left[
    \delta_{nm}
	\log \frac{r_n}{a}
    + 
    (1 - \delta_{nm}) 
    \log \frac{\Delta^{(p)}_{mn}}{a}
    \right].
    \nonumber
\end{align}
We used this matrix to 
{estimate} $k_p^\text{min}$ 
shown in Figures \ref{fig:freq_tuning} and \ref{fig:tunability}. 
{In these figures, black dashed lines are the} 
solutions of
\begin{equation}
    \det \mathbb{M}^{\text{($p$)}}(\boldsymbol{q}=\boldsymbol{0}) = 0
    \label{eq:det_cluster}
\end{equation}
since the minimal plasma frequency correspond{s} to 
a single cluster {of all wires at} 
the unit cell center
-- see Fig.~\ref{fig:deforms}.

The interaction between clusters can be approximated via $D(\boldsymbol{\Delta}_{lp})$, {where $\Delta_{lp}$ is the
vector between the corresponding cluster
centers. Thus,} 
each wire in the $l^\text{th}$ cluster influences every wire in the $p^\text{th}$ cluster equally, since $\Delta_{mn}^{(p,l)} \ll a,b,\Delta_{lp}$ in our approximation. 

The system (\ref{eq:system_n}) for clustered wire media in Fig.~\ref{fig:gen_cell_clusters} transforms in{to} a block 
matrix after re-indexing of wires:
\begin{align}
    &
	\left(
	\begin{array}{c|c|c}
        \mathbb{M}^\text{(1)}_{N_1\times N_1} \left( r_n\big|_{n=1}^{N_1} \right)
		& D(\boldsymbol{\Delta}_{21}) \, \mathbb{J}_{N_1\times N_2}
		&  \cdots \\
		\hline
		D(\boldsymbol{\Delta}_{12}) \, \mathbb{J}_{N_2\times N_1}& \mathbb{M}^\text{(2)}_{N_2\times N_2} \left( r_{N_1 + n} \big|_{n=1}^{N_2} \right)
		& \cdots \\
		\hline
		\vdots & \vdots&  \ddots  \\
	\end{array}
	\right)
	\cdot
    \nonumber \\
    & 
    \qquad \qquad \qquad \qquad
    \cdot
	\left(
	\begin{array}{c}
		\mathbf{I}^{(1)}_{N_1 \times 1} \\
		\hline
		\mathbf{I}^{(2)}_{N_2 \times 1} \\
		\hline
		\vdots
	\end{array}
	\right)
    = \boldsymbol{0},
    \label{eq:system_clusters}
\end{align}
where $\mathbb{J}_{N_p\times N_l}$ is {an $N_p\times N_l$ all-ones matrix, and} 
$N_l$ is {the number of wires in the} 
$l^\text{th}$ cluster. 

For example, 
when wires in 
{the considered} geometries 
approach a unit cell boundary, this clustering becomes clear. In Fig.~\ref{fig:geoms_lim}{,} these clusters are highlighted with different colors. The cluster 
approach allows 
{summing fewer} 
series for analysis of such systems and for {obtaining} 
$k_p^\text{lim}$ defined in the paper (see Fig.~\ref{fig:freq_tuning}).

Relative error of {analytically estimated} $k_p^\text{lim}$ 
(compared to numerically obtained results in COMSOL) decreases with increase of the $a/r_0$ ratio (with decrease of the wire 
radii){,} as can be concluded from Table \ref{tab:estimations_lim}.
{In any case,} 
the relative error for $k_p^\text{lim}$ and $k_p^\text{min}$ is much higher than for $k_p^\text{max}${,} provided in Tables \ref{tab:estimations_max_hex} and \ref{tab:estimations_max_sqr} ($\sim1\%$ versus $<0.001\%$ for very thin wires having $a/r_0=10^3$) even for small wire 
radii. {As the} 
wire 
radii {decrease, the} 
distances between all wires {also decrease, requiring calculations of $D$ near the singularity points to estimate $k_p^\text{min}$ and $k_p^\text{lim}$. To overcome this problem, we propose} 
approximations of the interaction constant $D$ {near the} 
singularity points 
in Eq.~{(B1)}

Hence, the main source of 
{the difference between} 
the 
{analytically estimated} (see Fig.~\ref{fig:tunability}) {and} 
the numerically obtained {tunability} values is the discrepancy {between} 
$k_p^\text{min}$ {estimated with} 
Eq.~(\ref{eq:det_cluster}) {and numerically estimated}.

\nocite{*}

\bibliography{apssamp}

\end{document}